\documentclass{aa}  
\usepackage{hyperref}
\hypersetup{colorlinks=true,
            allcolors=blue}
\usepackage{graphicx}
\usepackage{booktabs}
\usepackage{wasysym} 
\usepackage{ragged2e} 
\usepackage{multirow}
\usepackage{gensymb}

\newcommand{\kms}[0]{km\,s$^{-1}$}

\newcommand{\hei}[0]{\ion{He}{i}\,10833\,\AA\ }
\newcommand{\ha}[0]{\ion{H}{i}\,6565\,\AA\ }

\newcommand{\cplus}[0]{\mbox{CRIRES$^+$}}
\usepackage{txfonts}

\begin{document} 

\title{Probing the extent of WASP-52~b's atmosphere}
\subtitle{High-resolution observations and 3D modeling insights}

\author{F. Nail \inst{1},
        A. Oklop\v{c}i\'{c} \inst{1},
        M. MacLeod \inst{2},
        K. Baka\inst{3},
        S. Czesla\inst{4},
        E. Nagel\inst{5},
        D. Linssen\inst{1},
        J. Matthijsse\inst{6}
        }
    
\authorrunning{F. Nail et al.}
\institute{Anton Pannekoek Institute for Astronomy, University of Amsterdam, 1090 GE Amsterdam, Netherlands
\and 
Center for Astrophysics, Harvard \& Smithsonian, 60 Garden Street, MS-51, Cambridge, MA 02138, USA
\and 
Ludwig Maximilian University, Faculty of Physics, University Observatory, Scheinerstr. 1, Munich D-81679, Germany
\and
Thüringer Landessternwarte Tautenburg, Sternwarte 5, D-07778 Tautenburg, Germany
\and
Institut für Astrophysik und Geophysik, Georg-August-Universität, Friedrich Hund Platz 1, D-37077 Göttingen, Germany
\and
Planetary Exploration, Technical University Delft, Kluyverweg 1 2629 HS Delft, Netherlands
}
\date{Accepted for publication in Astronomy \& Astrophysics on August 22, 2025.}

\abstract{
WASP-52~b is an inflated hot Jupiter with a large Roche lobe filling fraction, positioned in the hot Neptune desert. Previous in-transit observations of the helium triplet at 10833~\AA\ have reported a range of excess absorption values (1.5\%–5.5\%) and a lack of net blueshift relative to the planet's rest frame, distinguishing it from other escaping atmospheres.  

This study investigates the extent and morphology of material escaping from WASP-52~b, assessing whether its outflow resembles a stream-like structure, as suggested for HAT-P-67~b and HAT-P-32~b. We obtained high-resolution spectra with CRIRES$^+$ and CARMENES, covering a broader orbital phase range ($\varphi \approx \pm0.1, \pm0.2, 0.5$) than previous studies. By analyzing the \hei line as a tracer of escape, we search for extended absorption beyond transit. Additionally, we explore possible outflow morphologies with three-dimensional (3D) hydrodynamic simulations, coupled with an improved radiative transfer approach, assessing the \hei triplet. 

The helium line shows no significant evidence of planetary material at the orbital phases observed in this work, though 3D modeling suggests such a structure could exist below observational detection limits.  

We conclude that the atmospheric outflow of WASP-52~b can be characterized by an intermediate hydrodynamic escape parameter, placing it in a transitional regime between cold outflows forming a stream-like morphology and hot outflows forming a tail. Additionally, the absence of a detectable in-transit blueshift in the helium line rules out a strong day-to-nightside anisotropy scenario. 
}

\maketitle

\section{Introduction}
WASP-52~b is a transiting exoplanet first identified in 2013 through the Wide Angle Search for Planets (WASP) survey \citep{hebrard_wasp-52b_2013}. This inflated hot Jupiter orbits a K-dwarf star of young to middle age (system parameters are listed in Table~\ref{ch4:tab:sysParam}). WASP-52~b is notable for its large size relative to its Roche Lobe, with $R_p / R_{\rm RL} \approx 0.58$, suggesting significant tidal interactions with its host star. In terms of surface gravity and Hill radius, the planet is similar to HAT-P-32~b \citep[see Fig.~9 in][]{macleod_streams_2024}. This makes WASP-52~b an intriguing candidate for investigating atmospheric escape in the regime of extended, stream-like outflows, similar to those observed in planets like HAT-P-67~b and HAT-P-32~b \citep{gully-santiago_large_2024, zhang_detection_2023}.

WASP-52~b has been the subject of numerous spectroscopic studies aiming to characterize its atmosphere \citep{kirk_transmission_2016, louden_precise_2017, mancini_orbital_2017, chen_gtc_2017, alam_hst_2018, bruno_starspot_2018, may_mopss_2018, kirk_kecknirspec_2022, allart_homogeneous_2023, canocchi_probing_2024, fournier-tondreau_transmission_2024}. Evidence of atmospheric escape has been obtained through detections in the helium triplet at 10833~\AA. Using ultra-narrowband photometry, \cite{vissapragada_constraints_2020} detected helium absorption of $0.29\% \pm 0.13\%$, which was later confirmed by Keck/NIRSPEC observations with a spectral resolution of R = 25~000. These observations revealed an excess absorption of up to 3.4\%, with a velocity shift relative to the planet's rest frame consistent with zero \citep{kirk_kecknirspec_2022}. 

The lack of a net blueshift is unusual compared to most exoplanets with detected helium signatures, where the line centroid typically shows a net blueshift corresponding to velocities of several \kms \citep[e.g.][]{nortmann_ground-based_2018, spake_helium_2018, allart_spectrally_2018, palle_he_2020, kirk_confirmation_2020, orell-miquel_confirmation_2023, zhang_detection_2023, masson_probing_2024}. This blueshift is generally interpreted as evidence of day-to-nightside winds in the planet's upper atmosphere, driven by pressure gradients between the strongly irradiated dayside and the cooler, shadowed nightside \citep[e.g.][]{nail_effects_2024}. Interestingly, HAT-P-67~b and HAT-P-32~b also show no significant velocity shift during transit, which aligns with the 3D hydrodynamic model predictions of \cite{nail_cold_2024}, that a cold stream-forming outflow should have an approximately zero velocity shift. This further motivates our investigation into whether WASP-52~b has a similar outflow structure.  

Additionally, \cite{allart_homogeneous_2023} observed the helium triplet using the high-resolution spectropolarimeter SPIRou (R = 70~000) on the Canada–France–Hawaii Telescope (CFHT), placing an upper limit of 1.69\% on the in-transit absorption. Furthermore, JWST NIRISS/SOSS observations (R = 700) by \cite{fournier-tondreau_transmission_2024} detected helium excess absorption with a significance of 7.3$\sigma$, corresponding to an estimated absorption of 5.5 $\pm$ 0.9\% in high-resolution observations. Additionally, they reported a tentative detection of post-transit absorption at the 2.9$\sigma$ level. Notably, all previously mentioned transit observations only covered orbital phases within a limited range of $-0.06$ to $0.07$ at most.

WASP-52 is an active star \citep{bruno_wasp-52b_2020, rosich_correcting_2020, salisbury_monitoring_2021}, with a $\log R'_{HK}$ index of approximately -4.4 \citep{hebrard_wasp-52b_2013}. It shows frequent star-spot crossings during planetary transits \citep[e.g.][]{mancini_orbital_2017, louden_precise_2017, may_mopss_2018, bruno_starspot_2018, bruno_wasp-52b_2020, fournier-tondreau_transmission_2024}. \cite{mancini_orbital_2017} accounted for star-spot occultation events during transit observations to derive a more precise sky-projected orbital obliquity of $\lambda_{\rm ob} = 3.8^{\circ} \pm 8.4^{\circ}$. This value remains broadly consistent with the earlier estimate from the Rossiter-McLaughlin effect, $\lambda_{\rm ob} = 24^{\circ}{}^{+17}_{-9}$, reported by \cite{hebrard_wasp-52b_2013}.

\begin{table}
\caption{System parameters of WASP-52~b from \cite{hebrard_wasp-52b_2013}.}
    \centering
    \begin{tabular}{llc}
    \toprule
    \toprule
        Parameter &  & Value \\
    \midrule
        Orbital period & $P_{\rm orb}$ [d] & $1.7497798\pm0.0000012$\\ 
        Transit midpoint & $T_c$ [d] & $2455793.68143\pm0.00009$ \\
        Transit duration & $T_{14}$ [h] & $1.8096\pm0.0120$ \\
        Planet mass & $M_p$ $[M_{\rm jup}]$ & $0.46 \pm 0.02$\\
        Planet radius & $R_p$ $[R_{\rm jup}]$& $1.27\pm0.03$ \\
        Eccentricity & $e$ & 0 \\
        Inclination & $i$ [deg] & $85.35\pm0.20$ \\
        Semi-major axis & $a$ [au] & $0.0272\pm0.0003$ \\
        Star mass & $M_*$ [$M_{\astrosun}$] & $0.87\pm0.03$ \\
        Star radius & $R_*$ [$R_{\astrosun}$]& $0.79\pm0.02$  \\
        Rotational period & $P_{\rm{rot}, *}$ [d] & $11.8\pm3.3$ \\
        Stellar Age & [Gyr] & $0.4^{+0.3(a)}_{-0.2}$\\
    \bottomrule
    \end{tabular}\\
    \small $^{(a)}$ The age may be underestimated; \cite{mancini_orbital_2017} estimated it could be up to 9.4 Gyr.
    
    \label{ch4:tab:sysParam}
\end{table}

WASP-52~b is located in a unique position within the hot Neptune desert, residing in the high-mass range with a short orbital period \citep{owen_photoevaporation_2018, vissapragada_constraints_2020}. This location makes it an intriguing case for studying hot Jupiter formation and evolution. \cite{owen_photoevaporation_2018} suggested that WASP-52~b may have reached its current position through a mechanism other than high-eccentricity migration, or that its inflation occurred only after the planet had fully circularized.

WASP-52~b is a compelling candidate for studying the escape of atmospheric material due to its location in the hot Neptunian desert, its significant Roche lobe filling, and the observed absence of a blueshift in helium absorption. We have conducted observations using the CARMENES spectrograph at the Calar Alto Observatory and the \cplus spectrograph at the Very Large Telescope (VLT) in the visible (VIS) and near-infrared (NIR). These observations cover a larger phase range of the planet's orbit than usual to constrain the full extent of escaping planetary material. By combining these observations with 3D hydrodynamic models, we aim to constrain the atmospheric outflow morphology. Specifically, we investigate whether WASP-52~b has a widely extended outflow similar to those observed in HAT-P-67~b and HAT-P-32~b \citep{gully-santiago_large_2024, zhang_detection_2023}, which may be associated with a stream morphology \citep[][]{nail_cold_2024, macleod_streams_2024}. In addition, we introduce a new and improved approach for performing radiative transfer analysis on our 3D simulations. The new method is based on the use of the \texttt{Cloudy} code \citep{ferland_cloudy_1998}. Cloudy's extensive database of reaction rates and cross-sections makes this method more appropriate for simulating low-temperature outflows that produce stream-like morphologies, as indicated by our previous work \citep{nail_cold_2024}. Additionally, \texttt{Cloudy} calculations include dozens of atomic and ionic species, which can potentially support investigations of additional tracers of atmospheric escape \citep[e.g.][]{linssen_expanding_2023-1} in the context of large-scale 3D simulations.

This study is organized as follows: In Section \ref{ch4:sec:Results_observations}, we describe our observing strategy, the data obtained for this study, the data reduction process, and we present our observational results from the helium line. In Section \ref{ch4:sec:Results_3DSimulations}, we describe the 3D simulation setup with \texttt{Athena++}, the new framework developed for post-processing to generate synthetic spectra from these models, and we present the 3D modeling results and compare them to the observational helium data. This is followed by a discussion in Section \ref{ch4:sec:discussion} and a brief summary in Section \ref{ch4:sec:summary}.

\section{\hei observations} \label{ch4:sec:Results_observations}

We obtained high-resolution spectra of WASP-52 in the visible and near-infrared, covering multiple orbital phases outside of transit, to investigate the extent of atmospheric escape of planet WASP-52~b. The metastable helium line \hei is a commonly used tracer for atmospheric escape in low-density regions \citep[e.g.][]{oklopcic_new_2018, nortmann_ground-based_2018, allart_homogeneous_2023, ballabio_understanding_2025}, making it an ideal starting point for our analysis. Additional potential spectral tracers of atmospheric escape, particularly those detectable with a high-resolution spectrograph in the visible wavelength range, may also provide valuable insights into the composition and dynamics of the outflow. However, exploring such tracers is beyond the scope of this study; here, we focus exclusively on helium.

\begin{figure}[h]
    \centering
    \includegraphics[width=0.4\textwidth]{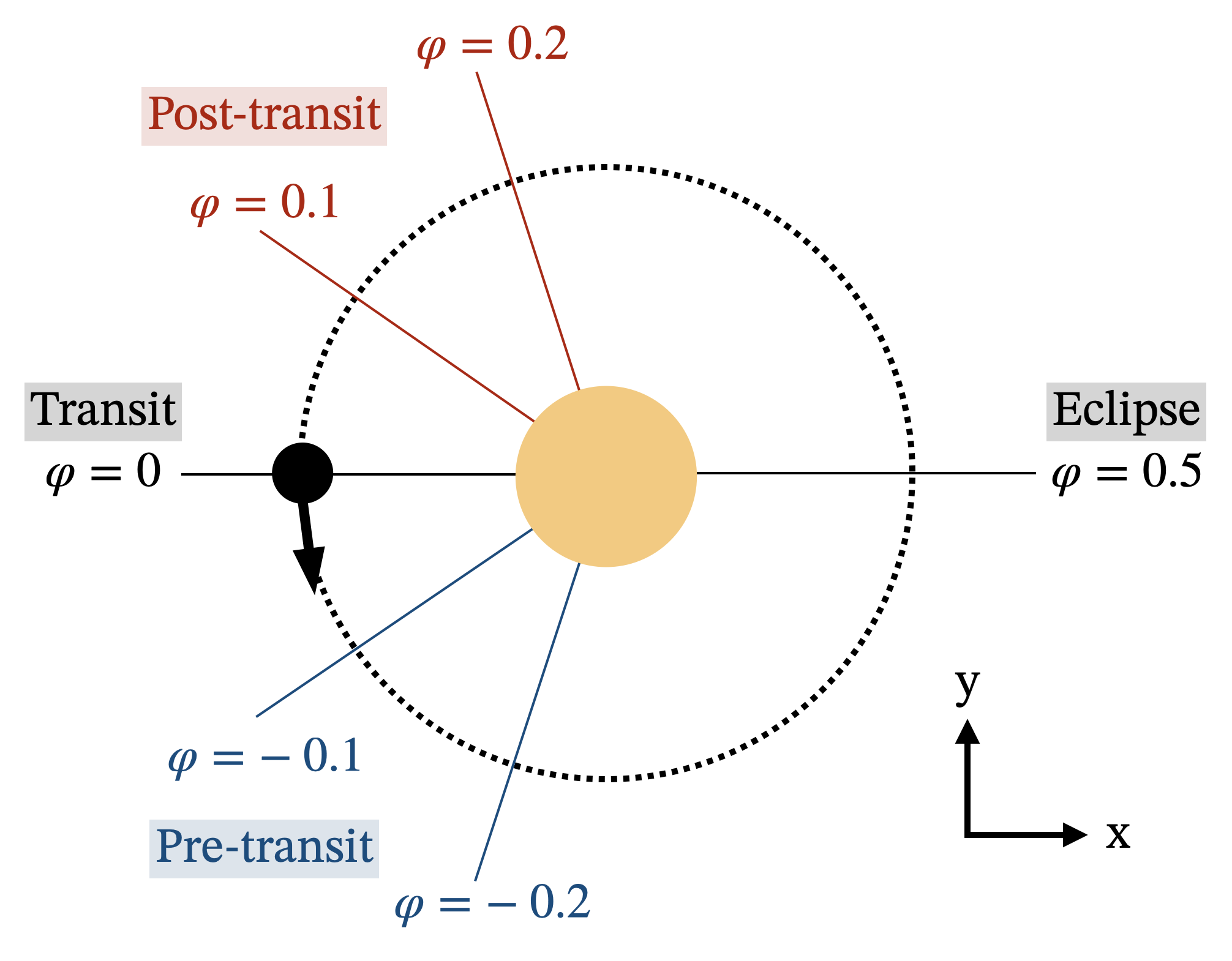}
    \caption{Illustration (not to scale) of the observing strategy for the WASP-52 system. The planet orbits the central star counterclockwise, with different phase angles, $\varphi$, representing the observer's line of sight to the system in the orbital mid-plane. During negative phase angles (pre-transit), the observer probes a potential leading tail of escaping atmosphere, while positive phase angles (post-transit) could reveal additional absorption from a trailing tail. Observations were conducted during both pre- and post-transit phases to assess the extent of atmospheric escape. For an out-of-transit spectrum, where the planet’s atmospheric contribution is expected to be minimal, additional observations were taken near eclipse, when the planet is obscured by the star.}
    \label{ch4:fig:sketch}
\end{figure}

Figure \ref{ch4:fig:sketch} illustrates our observing strategy. Unlike traditional transit observations, where a time series of spectra is collected throughout the transit with a baseline of a few hours, we targeted individual phases independently to detect any excess absorption compared to spectra taken near eclipse, when the absorption signal from the planetary outflow is expected to be minimal. We did not gather in-transit spectra, as these observations have already been conducted in previous studies. Instead, we focused solely on the pre- and post-transit phases at $\varphi \approx\pm 0.1$ and $\varphi \approx\pm 0.2$, allowing for a margin of $\pm 0.05$ to accommodate observational constraints.

\subsection{\cplus data}
We obtained spectra of WASP-52 using the high-resolution spectrograph \cplus\ \citep{dorn_crires_2023}, mounted on the Very Large Telescope (VLT). Utilizing a $0.2''$ slit, we achieved a nominal spectral resolution of approximately R = $100~000$. To effectively subtract the sky background and reduce instrumental effects, we applied an ABBA nodding pattern, alternating the target between two distinct slit positions (A and B). 

The observing nights and according orbital phases are listed in Table \ref{ch4:tab:CRIRES_obs}. On night 1, phases $-0.2$ and $-0.1$ were observed consecutively, yielding a total of 8 spectra (4 x AB nodding). On nights 2 and 3, two exposures were taken at each nodding position, resulting in 4 spectra per night (2 x AB nodding). Unfortunately, phase 0.1 could not be observed due to poor weather conditions. Each spectrum had an integration time of 900 seconds and was taken in the Y-band using the Y1029 wavelength setting.

\begin{table}
\caption{Overview of observation nights with \cplus.}
    \centering
    \begin{tabular}{ccccc}
    \toprule
    \toprule
    Night & Date & $\overline{\varphi}$ & \# Spectra \\
    \midrule
    1 & 30/09/2023 & $-0.17$ & 4\\
      &            & $-0.14$ & 4\\
    2 & 08/10/2023 & $+0.42$ & 4 \\
    3 & 13/10/2023 & $+0.27$ & 4 \\
    \bottomrule
    \end{tabular}\\
    \small Notes. $\overline{\varphi}$ represents the central orbital phase.
    \label{ch4:tab:CRIRES_obs}
\end{table}

We used the \cplus\ pipeline, {\tt cr2res}, provided by ESO\footnote{\url{https://www.eso.org/sci/software/pipelines/index.html\#pipelines_table}} to reduce the data. The raw calibration frames, acquired as part of the observatory's daily routine, were processed through the standard calibration cascade outlined in the \cplus\ Pipeline User Manual. This step generated the processed calibration files, including the bad pixel mask, normalized flat field, and the trace-wave file containing the wavelength solution and spectral order positions. Next, we reduced the raw science frames with the \verb!cr2res_obs_nodding! recipe. This step involved applying the processed calibration files to the science frames, subtracting the B frames from the A frames in each AB (or BA) nodding sequence, and performing optimal extraction to produce one-dimensional science spectra. We then applied the wavelength solution to the extracted spectra.

\begin{figure}
    \centering
    \includegraphics[width=0.5\textwidth]{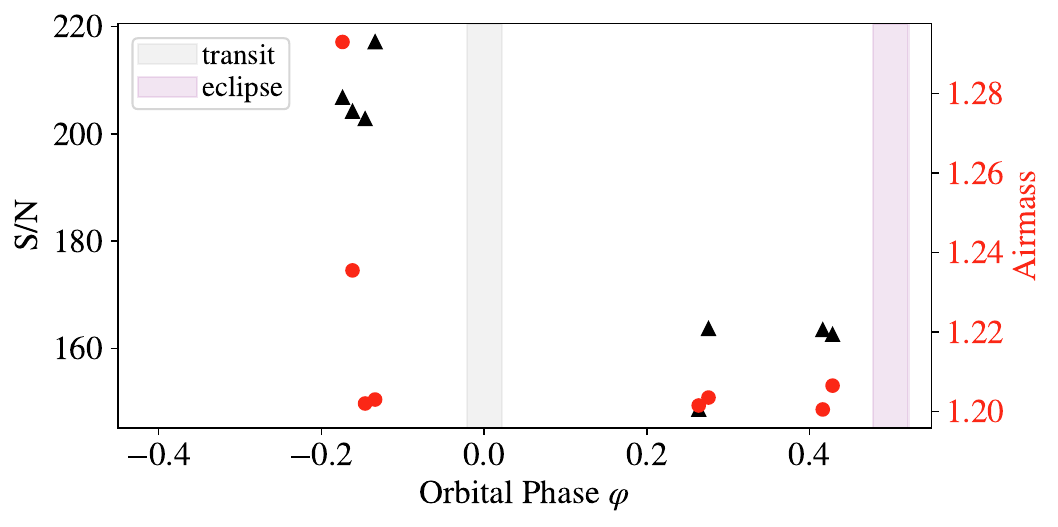}
    \caption{Air mass (red) and signal-to-noise ratio (black) in the region of the \hei line ($10820.0~\text{\AA} < \lambda< 10870.1~\text{\AA}$) of the \cplus observations. Shaded regions indicate the start and end of the planet's transit and eclipse. }
    \label{ch4:fig:airmass_SN_cplus}
\end{figure}

To enhance the signal-to-noise ratio, we combined two consecutive spectra from each nodding position. The resulting spectrum was then corrected for telluric absorption lines using \texttt{Molecfit} \citep{smette_molecfit_2015, kausch_molecfit_2015}. Figure~\ref{ch4:fig:airmass_SN_cplus} shows the air mass evolution alongside the mean signal-to-noise ratio (S/N) near the \hei line in the telluric-corrected spectra.

\subsection{CARMENES data}
Similarly, we collected spectra of WASP-52 using the CARMENES spectrograph \citep{quirrenbach_carmenes_2018}, mounted on the 3.5~m telescope at the Calar Alto Observatory. CARMENES provides simultaneous coverage of both the visual ($5500-9600$~\AA) and near-infrared ($9600-17200$~\AA) regions, with nominal spectral resolutions of R=$94~600$ and $80~400$, respectively. During the observation campaign, each spectrum was obtained with an exposure time of 900 seconds. The observing nights are listed in Table \ref{ch4:tab:CARMENES_obs}.

\begin{figure}[h]
    \centering
\includegraphics[width=0.5\textwidth]{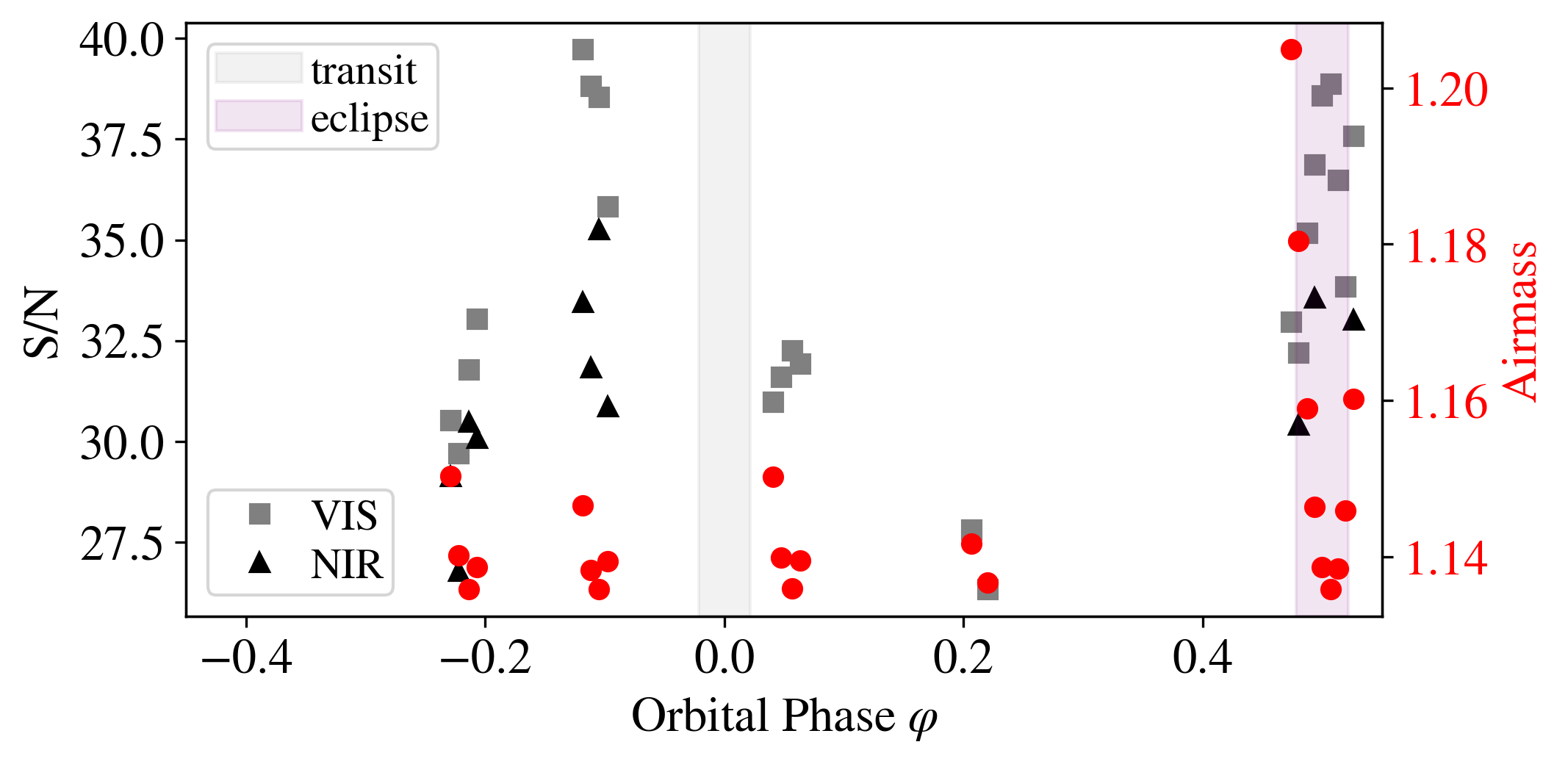}
    \caption{Air mass and signal-to-noise ratios (S/N) of the CARMENES observations (combined spectra). Black triangles indicate the S/N in the region of the helium triplet ($10820.0$~\AA$ ~< \lambda< 10870.1$~\AA), while the gray squares represent the S/N in the region of the \ha line ($6550.0$~\AA~$ < \lambda< 6580.0$~\AA). The red dots show the air mass during the observation.}
    \label{ch4:fig:airmass_SN_carm}
\end{figure}

We processed the raw spectra using the CARMENES pipeline CARACAL \citep{zechmeister_flat-relative_2014, caballero_carmenes_2016}, which performs standard data reduction steps, including dark subtraction, flat fielding, wavelength calibration, and spectrum extraction. The pipeline also provides noise estimates for each data point. To enhance data quality, wavelength points with low signal-to-noise ratios (S/N) were masked. We used {\tt Molecfit} to performed the telluric correction. Figure \ref{ch4:fig:airmass_SN_carm} shows the air mass and S/N near the helium line in the near-infrared, along with the S/N near the \ha line for the reduced spectra. 

\begin{table}
\caption{Overview of observation nights with CARMENES.}
    \centering
    \begin{tabular}{cccccc}
    \toprule
    \toprule
    \multirow{2}{*}{Night} & \multirow{2}{*}{Date} & \multirow{2}{*}{$\overline{\varphi}$} & \multicolumn{2}{c}{\# Spectra} \\
    &  &  & VIS & NIR \\
    \midrule
    1 & 05/09/2023 & $+0.50$ & 9 & 3 \\
    2 & 08/09/2023 & $+0.21$ & 2 & 0 \\
    3 & 09/09/2023 & $-0.22$ & 4 & 4 \\
    4 & 20/09/2023 & $+0.05$ & 4 & (4)$^{*}$ \\
    5 & 02/10/2023 & $-0.11$ & 4 & 4 \\
    \bottomrule
    \end{tabular}\\
    \small Notes. $^{*}$Readout error for NIR images; spectra not used in this work.
    \label{ch4:tab:CARMENES_obs}
\end{table}

\subsection{Searching for planetary excess absorption in the helium line core}
Since our observing strategy does not permit the creation of a high signal-to-noise master-out-of-transit spectrum, typically constructed by co-adding multiple out-of-transit spectra, we instead focus on line core equivalent widths (EW) of the stellar \hei line. Studies by \cite{gully-santiago_large_2024} and \cite{zhang_giant_2023} show that changes in the stellar equivalent width caused by planetary excess absorption are clearly detectable in their respective light curves (see their Fig.~9 and Fig.~1).

\begin{figure}
    \centering
    \includegraphics[trim={0 1cm 0 0},clip,width=0.5\textwidth]{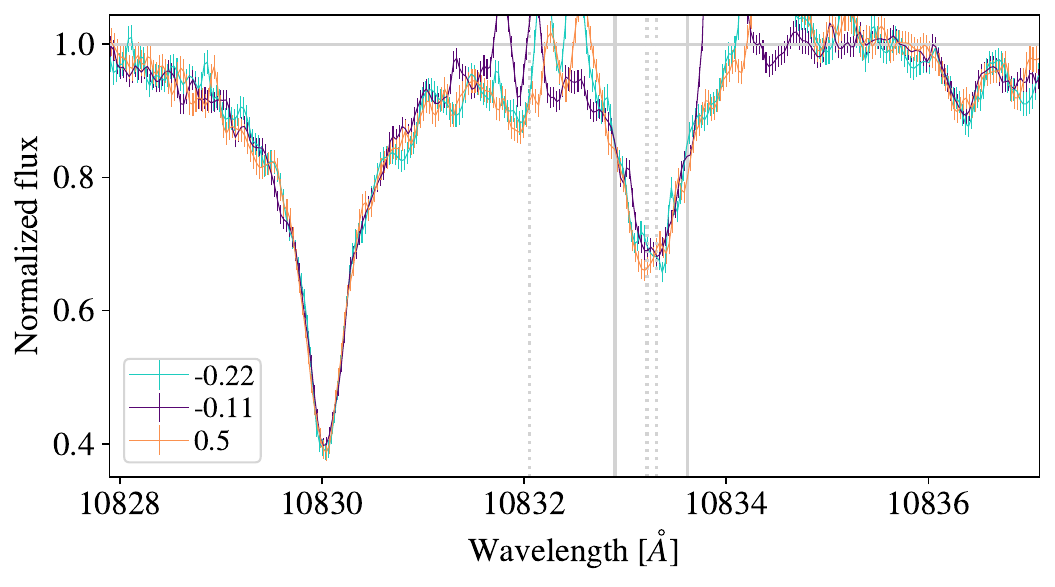}
    \includegraphics[width=0.5\textwidth]{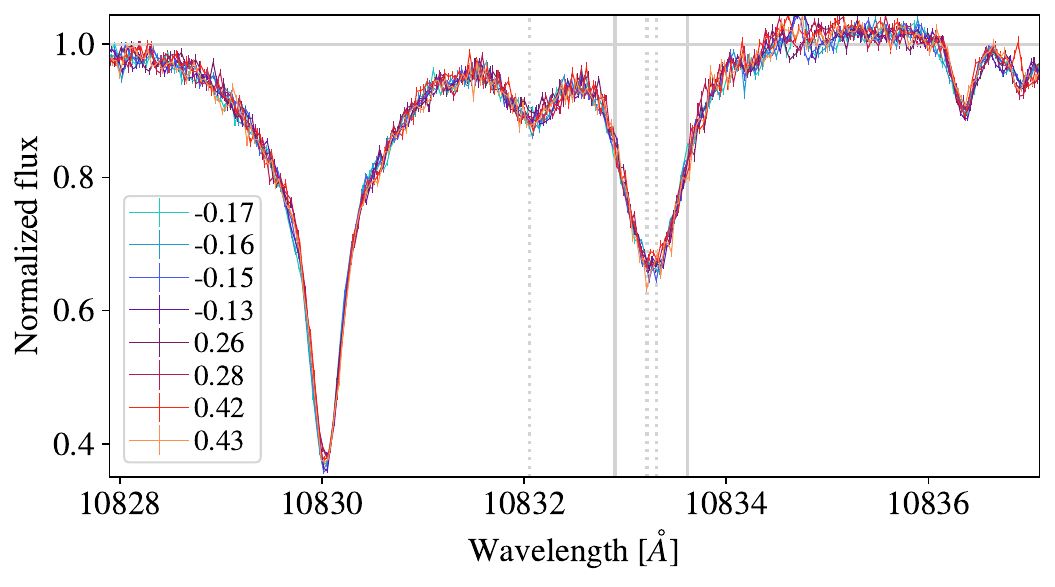}
    \includegraphics[width=0.5\textwidth]{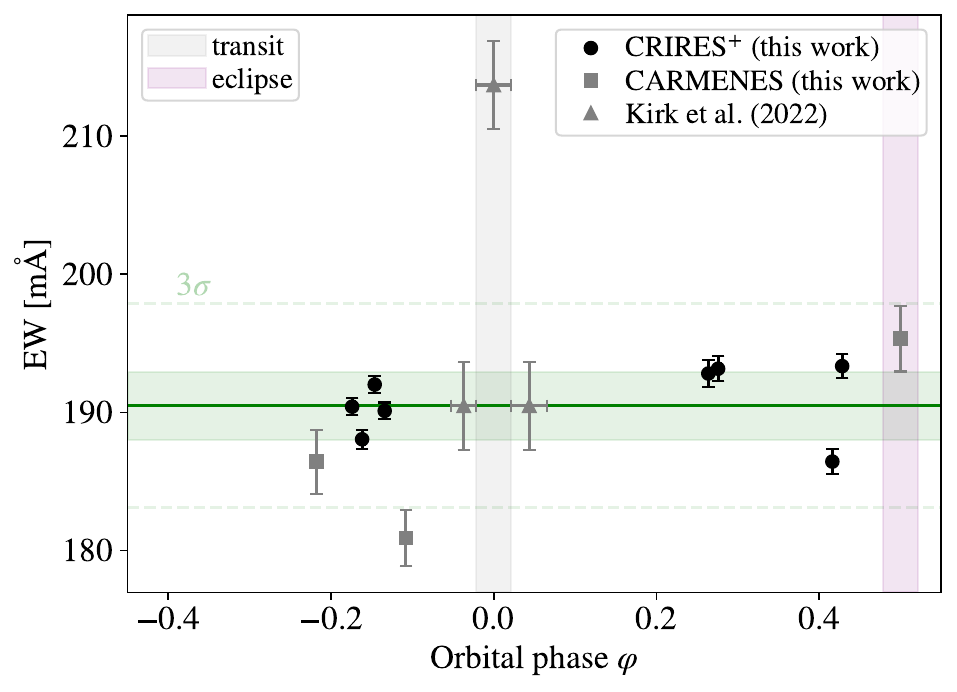}
    \caption{Stellar \hei line observed with CARMENES (top), \cplus\ (middle), and the equivalent widths of the line core across orbital phases (bottom). Dashed lines in the top and middle panels indicate the vacuum wavelengths of the helium triplet, while solid lines mark the line core region used for EW calculations. Observations from this study (squares and dots in the bottom panel) show no significant variability in the stellar helium triplet. We interpret the data points as a baseline of spectra uncontaminated by planetary absorption, yielding a weighted mean EW of $(190.5 \pm 2.5)$~m\AA\ (green line). For reference, we include the estimated mid-transit equivalent width based on the observations by \citet{kirk_kecknirspec_2022}, obtained by adding the reported planetary excess to the stellar baseline measured from our data.}
    \label{ch4:fig:Helium_obs}
\end{figure}

In Figure~\ref{ch4:fig:Helium_obs}, we present the stellar metastable helium triplet as observed with CARMENES and \cplus. CRIRES$^+$ at the VLT uses AB nodding, a technique that effectively eliminates telluric emission features. This method alternates between two positions (A and B) on the sky, capturing both the target spectrum and telluric emission lines. By subtracting the two exposures (A – B), the telluric emission, which is uniform across the slit, cancels out, leaving a cleaner target spectrum. In contrast, CARMENES data still contains telluric emission lines, as it does not use this technique. Hence, while the CARMENES spectra shown (top panel) were corrected for telluric absorption lines with \texttt{Molecfit}, telluric emission lines remain. We chose not to model and subtract the telluric OH emission lines to avoid potential inaccuracies. Instead, we carefully scheduled the observations to ensure that the OH emission lines do not overlap with the core of the helium main component. As such, we focus exclusively on the line core ($\lambda = 10832.90 - 10833.62\,$\AA) of the main component, highlighted by the solid vertical lines. We chose this wavelength range in an effort to minimize interference from OH emission lines, though some effect may still remain. Furthermore, \citet{gully-santiago_large_2024} and \citet{zhang_giant_2023}, in their observational studies of extended atmospheric escape, found that the planetary absorption signal remains approximately stationary in the stellar rest frame (see their Fig. 10), exhibiting little correlation with the planet's orbital motion. Based on these findings, we are confident in restricting our analysis to the stellar line core, which is centered in the stellar rest frame. In Appendix~\ref{app:full_EW}, we present an additional analysis of the full helium line using only the \cplus\ data, which allows full line coverage due to its corrections for both telluric absorption and emission lines. The qualitative results align with those obtained from the line core analysis presented below.

In the bottom panel of Figure~\ref{ch4:fig:Helium_obs}, we present the core equivalent width derived from the helium spectra of both data sets, integrating the flux exclusively over the wavelength range corresponding to the helium line core, as indicated. 

Furthermore, we use the results by \cite{kirk_kecknirspec_2022} (Keck/NIRSPEC) as a reference for planetary excess helium absorption. Using \texttt{WebPlotDigitizer}\footnote{\url{https://github.com/automeris-io/WebPlotDigitizer}}, we extracted the helium line profile of WASP-52~b presented in their work (see their Fig.~7). Since the variability of the stellar line is accounted for by dividing by the master-out-of-transit spectrum, the derived excess absorption is independent of stellar variability. From the extracted line profile, we measured the equivalent width of the helium line core within the same wavelength ranges used for our observations, resulting in an excess absorption core EW of $(23 \pm 2)$~m\AA. The uncertainty in this measurement reflects an estimated error margin, as the direct error derived from the data points' standard deviation is only $1.14$~m\AA. However, due to reliance on \texttt{WebPlotDigitizer} for data extraction and the sensitivity of the EW to data binning, we adopted a more conservative error estimate. We interpreted the pre- and post-transit absorption signals reported by \citet{kirk_kecknirspec_2022} to be consistent with zero within the estimated uncertainties. For the out-of-transit phases, we assumed the same noise level and uncertainties as during mid-transit, using the same error value of 2~m\AA. In the bottom panel of Figure~\ref{ch4:fig:Helium_obs}, we added the results from \cite{kirk_kecknirspec_2022} to the baseline of our observations, which itself carries some uncertainty. Consequently, the error bars of the Keck observations represent the combined uncertainties from both the absorption measurement and the standard deviation of the baseline.

From our observations, the EW light curve of WASP-52~b shows no evidence of excess absorption outside the transit, with the measured data points clustering closely around a consistent EW value. Therefore, we interpret all our data points as reflecting a baseline of stellar absorption, with no apparent contribution from the planet. Notably, there is one outlier with lower absorption at phase $-0.11$, which corresponds to the CARMENES spectrum with the strongest OH emission line. We interpret this as the likely cause of the deviation.  

To account for the different uncertainties in each data point, we calculated the weighted mean and weighted standard deviation of the equivalent widths using inverse-variance weighting. Each data point was assigned a weight proportional to the inverse square of its associated uncertainty, ensuring that points with smaller errors contributed more significantly to the results. The weighted mean was computed as the sum of the product of the weights and the equivalent widths, divided by the total weight. The weighted standard deviation was then determined by measuring the weighted spread of the data around the weighted mean. We derive a stellar helium line core EW of 194.5~m\AA, with a weighted standard deviation of 2.5~m\AA. Our data sensitivity allows us to detect changes in the equivalent width (EW) of the line core as small as 7.5~m\AA\ with 3$\sigma$ confidence. This level of precision is sufficient to detect the in-transit excess EW reported by \cite{kirk_kecknirspec_2022}. 

Unfortunately, due to bad weather conditions and technical difficulties, we lack post-transit data at the phase of 0.1 in the near-infrared, which would have been instrumental in exploring potential trailing gas from the planet. Based on the absence of any pre-transit excess absorption, we rule out the presence of a significantly extended leading helium stream like that observed in  HAT-P-67~b and HAT-P-32~b. Our observations constrain the extent of the absorbing gas to $31~R_p$ ($\varphi = -0.11$) in the leading direction and $73~R_p$ ($\varphi = 0.26$) in the trailing direction.

\section{3D simulations with \texttt{Athena++}} \label{ch4:sec:Results_3DSimulations}

\subsection{Simulation setup and models} \label{ch4:sec:Athena_methods}
We simulate a planet in a circular orbit around its host star, incorporating thermal winds from both the planet and the star, using the 3D Eulerian (magneto)hydrodynamic code \texttt{Athena++}\footnote{\url{https://github.com/PrincetonUniversity/athena}}, version 2021 \citep{stone_athena_2020}. For these hydrodynamic simulations, we use the setup previously developed in \cite{macleod_stellar_2022} and extended to account for planetary day-night anisotropies in \cite{nail_effects_2024}. In the following, we outline the most important features of our simulations and the modifications made for this study, building on the methods described in Section 2 of those works. In this approach, we remain agnostic about the detailed physical mechanisms driving the winds. This approach allows us to flexibly explore a broader parameter space. We apply a boundary condition that heats the base of the outflow, and approximate the gas temperature using a nearly isothermal equation of state. This simplification enables us to focus on the large-scale interactions between the planetary and stellar winds.

Within the \texttt{Athena++} simulations, the equations for mass, momentum, and energy conservation are solved for an inviscid gas. We use the ideal gas law with an adiabatic index of $\gamma = 1.0001$, which approximates isothermal behavior along adiabats. Nonetheless, temperatures at the points where planetary winds originate can vary, enabling us to account for temperature differences between the planet's dayside and nightside, as outlined in \cite{nail_effects_2024}. The simulation includes the gravitational influences of the planet and the star, while neglecting the outflow's back-reaction on the planetary orbit and the effect of \hei radiation pressure on gas dynamics.

Hydrodynamic winds are introduced by parameterizing the conditions at the bases of the stellar and planetary winds. The stellar surface pressure is determined by the constant surface density $\rho_*$ and the hydrodynamic escape parameter $\lambda_*$, which represents the ratio of gravitational potential energy per particle, $G M_* m R_*^{-1}$, to thermal energy, $k_{\rm B} T(R_*)$, where $m$ is the mass of the escaping particle. For all models, we set $\lambda_* = 15$. Similarly, the pressure at the planetary surface is given by 
\begin{equation}
    P_p = \rho_p \frac{GM_p}{\gamma \lambda_p R_p}~,
\end{equation}
where $\lambda_p = \frac{GM_p}{c_s^2 R_p}$ denotes the planet's hydrodynamic escape parameter. In models incorporating day-night anisotropy, surface pressure may vary across the planetary surface \citep{nail_effects_2024}.

We use a spherical polar mesh centered on the star in a co-rotating frame aligned with the planet's orbit. The computational domain spans from the star's surface out to a distance of 0.37\,au in all directions. The planet is positioned along the negative x-axis at a distance of $a = 0.027$\,au from the star. Both the star and the planet rotate with the planet's orbital frequency, $\Omega = \Omega_{\text{orb}} = G(M_p + M_*)a^{-3}$, with their angular momenta oriented in the positive z-direction. 

The base mesh consists of $9 \times 6 \times 12$ mesh blocks, each containing $16^3$ zones that are logarithmically spaced in the radial ($r$) direction and evenly spaced in the angular ($\theta$ and $\varphi$) directions, ensuring nearly cubic zone shapes. Near the poles, we mitigate extreme aspect ratios by averaging conserved quantities, effectively reducing the number of zones in the $\varphi$ direction.

To accurately resolve the atmospheric scale height near the planet, we used a mesh scaled to the planetary radius, setting the static mesh refinement (SMR) level to $N_{\mathrm{SMR}} = 3$ ensuring increased resolution within a radius of 12~$R_p$. We analyze snapshots of the simulation only after a quasi-steady state had been reached after five orbits, defined as the point where the planetary and stellar mass-loss rates stabilized. These rates are calculated by tracking changes in the planet's and star's mass over time while accounting for mass flux across the simulation boundaries. This approach ensures mass conservation by incorporating ejected, accreted, and redistributed mass within the system.

We investigate four models to explore the influence of stellar and planetary wind properties on the outflow morphology of WASP-52~b (see Tab.~\ref{ch4:tab:model_overview}):
\begin{itemize}
    \item[$\bullet$] ISO (Isotropic planetary wind): A spherically symmetric planetary outflow, interacting with a moderate stellar wind ($\dot{m}_* /  \dot{m}_p \approx 5$).
    \item[$\bullet$] ANISO (Anisotropic planetary wind): Same as ISO, but accounting for a day-night pressure gradient following the approach in \cite{nail_effects_2024}, with the nightside pressure reduced to 10\% of the dayside value.
    \item[$\bullet$] TORUS (Reduced stellar wind): Reduced stellar mass-loss rate relative to model ISO ($\dot{m}_* / \dot{m}_p \approx 0.5$).
    \item[$\bullet$] TAIL (Enhanced stellar wind): Enhanced stellar mass-loss rate relative to the ISO model ($\dot{m}_* / \dot{m}_p \approx 16$).
\end{itemize}

\begin{table*}
    \caption{Input parameters of the 3D simulation models.}
    \label{ch4:tab:model_overview}
     \centering
    \begin{tabular}{lcccccc}
    \toprule
    \toprule
         model & $\lambda_p$ & $f_{\rm pres}$ &  $\rho_p$ & $\rho_*$ & $c_{\rm s, sub}$  & $c_{\rm s, anti}$ \\ 
          & &  [\%] & [g~cm$^{-3}$] & [g~cm$^{-3}$] & [km~s$^{-1}$] & [km~s$^{-1}$] \\  
    \midrule
         ISO & 5 & 100 & $5.42 \times 10^{-16}$ & $2.87\times 10^{-14}$ & 11.3 & 11.3  \\
         ANISO & 5 & 10 & $1.81 \times 10^{-15}$ & $2.87\times 10^{-14}$ & 11.3 & 1.1 \\
         TORUS & 5 & 100 & $5.42 \times 10^{-16}$ & $2.87\times 10^{-15}$ & 11.3 & 11.3 \\
         TAIL & 5 & 100 & $5.42 \times 10^{-16}$ & $7.22 \times 10^{-14}$ & 11.3 & 11.3  \\
    \bottomrule
        \end{tabular}
    \\ \small Notes. $f_{\rm pres}$ is the pressure fraction at the antistellar point relative to the substellar point. \\At fixed $\mu = 0.6$, $c_s = 11.3$~\kms\ corresponds to $T=9300$~K, and $c_s = 1.1$~\kms\ to $T=90$~K.
\end{table*}

\subsection{Post-processing of 3D simulations} \label{ch4:sec:postprocessing_Cloudy}
To post-process the \texttt{Athena++} simulation snapshots, we introduce a new framework that connects \texttt{Athena++} snapshot outputs (3D pressure and density profiles) with \texttt{Cloudy}\footnote{\href{https://gitlab.nublado.org/cloudy/cloudy/-/wikis/home}{gitlab.nublado.org/cloudy/cloudy}} \texttt{v23.01} \citep{ferland_cloudy_1998, ferland_2017_2017, chatzikos_2023_2023, gunasekera_2301_2023}, a non-local thermodynamic equilibrium (NLTE) plasma simulation code.

Using atmospheric density and temperature profiles from \texttt{Athena++} and an incoming stellar spectral energy distribution (SED), \texttt{Cloudy} determines the electron temperature, ionization levels, and excitation states of various atomic species. Although \texttt{Cloudy} usually computes the electron temperature, it can also accept a predefined temperature profile, as is done here using data from the \texttt{Athena++} simulation. In reality, photoionization by stellar EUV/XUV photons can locally heat the gas above the parameterized temperatures, particularly in low-density regions. However, we chose the \texttt{Athena++} temperature profiles to maintain consistency with the hydrodynamic evolution of the wind. Our approach offers a computationally efficient compromise, combining realistic 3D dynamics with detailed 1D radiative transfer, though it neglects lateral radiative effects and self-consistent temperature evolution.

To simulate transmission spectra originating from orbital phases similar to the observations (see Fig.~\ref{ch4:fig:sketch}), one-dimensional rays are generated for the corresponding observing angles along the line-of-sight between the observer and the star. The code first calculates a direction vector based on the azimuthal and polar angles, then determines the number of rays required to cover the stellar surface based on the observing angle and position of the planet. To ensure higher resolution near the planet, rays are allocated more densely in that region, with their weighting in the spectra being adjusted according to their relative area, following the approach of the previous radiative transfer analysis described in \cite{macleod_stellar_2022}. The radial extent of each ray is set by user-defined inner and outer limits (here from -20 to 40~$R_p$), as well as the mesh grid containing the ray. To assign values to the zone centers, we use nearest-neighbor interpolation from the simulation grid, ensuring that any point within a zone's volume receives the corresponding value of that zone.

Density and pressure data from these rays, as well as the derived temperature 
\begin{equation}
T = \gamma \frac{\mu m_H }{k_{\rm B}} \frac{P}{\rho},
\label{ch4:eq:T}
\end{equation}
assuming a mean molecular weight of $\mu = 0.6$, were extracted from \texttt{Athena++} and provided to \texttt{Cloudy}. 

Since \texttt{Cloudy} performs radiative transfer calculations, it does produce a spectrum that is transmitted through each ray. However, since it is a hydrostatic code, this spectrum does not include spectral line broadening due to the radial velocities. Therefore, instead of using \texttt{Cloudy}'s own predicted spectrum, we instead extract from it the number densities of each energy level of each different element, and use these to calculate the transit spectrum in post-processing. In this way, we are able to include Doppler shifts and broadening of the spectral lines based on the velocity structure derived from \texttt{Athena++}.

Using the number densities of the energy levels for each ray as calculated by \texttt{Cloudy}, assuming solar composition, we performed radiative transfer calculations for the metastable helium triplet. For the radiative transfer, spectral line data for helium were sourced from the NIST database\footnote{\url{www.nist.gov/pml/atomic-spectra-database}}, such as wavelength, energy levels, oscillator strengths, and Einstein coefficients. 

The optical depth $\tau$ at each frequency $\nu$ was calculated along the path length d$s$ of the rays:
\begin{equation}
    \tau_{\nu} = \int n_s \sigma_0 \Phi(\Delta \nu)~\rm{d} s.
    \label{ch4:eq:tau}
\end{equation}
Equation \ref{ch4:eq:tau} takes into account the species number density $n_s$, the line cross-section $\sigma_0$, and the Voigt $\Phi(\Delta v)$ profile, which includes a Lorentzian component due to natural line broadening and a Gaussian component that depends on the local gas temperature $T$. Additionally, the wavelength shift $\Delta \nu$ reflects the gas velocity in the inertial frame, projected along the line of sight. The extinction of each ray is calculated by $1 -\exp(-\tau_{\nu})$. The rays for a given observing angle are weighted according to the quadratic limb darkening law, and their extinction contributions are summed to compute the total transmission spectrum. We tested the convergence of the spectra by varying the total number of rays used in the radiative transfer calculation, finding that approximately 1000 rays are sufficient to achieve reliable results (see Appendix~\ref{app:convergence}).

A spectral energy distribution (SED) for WASP-52 is unavailable, thus we adopted the SED of a comparable star, HD~40307, which has a spectral type of K2.5 and $T_{\rm eff}=4880$~K, provided by the MUSCLES Treasury Survey \citep{france_muscles_2016}, and scaled it to the orbital distance of WASP-52~b. The SED serves as a close approximation to the reported properties of WASP-52~b's host star, which has a spectral type of K2.0 and $T_{\rm eff}=5000$~K \citep{hebrard_wasp-52b_2013}. The age of HD~40307 is estimated to be 2.6 Gyr \citep{Turnbull_nearby_2015}, while \cite{hebrard_wasp-52b_2013} estimated WASP-52 to be 0.4 Gyr. However, WASP-52 is an active star, and its actual age remains uncertain, leading to potential differences in their XUV fluxes. If WASP-52 is younger and more active, its XUV flux would be higher; conversely, if it is older, as suggested by \cite{mancini_orbital_2017}, its XUV flux would be lower. These variations could affect the absolute populations of metastable helium in the gas to some extent.  

To assess the performance of the new radiative transfer framework, we compared its results to those obtained with the previous code introduced by \cite{macleod_stellar_2022}, which focuses solely on the \hei line, presented in Appendix \ref{ch4:app:old_new_RT_compare}. We found that the main difference between the codes is the inclusion of doubly ionized helium in \texttt{Cloudy}. The production of this species significantly reduces the metastable helium number density, as it is primarily produced through the recombination of singly ionized helium. In \texttt{Cloudy}, less singly ionized helium is available due to the new He$^{2+}$ pathway. The previous code uses fixed values of recombination rates and collision cross-sections that are well suited for modeling planetary outflows of temperatures around $10^{4}$~K, while \texttt{Cloudy} incorporates more accurate values across a broader temperature range. Therefore, the new radiative transfer framework is better suited for calculating abundances in large-scale 3D simulations of atmospheric escape, especially those aimed at studying stream-like outflows, as they might include a wide range of densities and temperatures a factor of a few lower than $10^{4}$~K. In Figure \ref{ch4:fig:rt_comparison}, we compare the helium and hydrogen species number densities along rays generated by the different codes. The largest differences are observed in low-density environments, where a significant population of He$^{2+}$ can build up due to the low probability of recombination. 

Additionally, we assess the physical timescales relevant to the population and observability of helium in the metastable triplet state. Based on \texttt{Cloudy} output, we estimate the total photoionization timescale of $\sim 1$ hour for neutral helium. The radiative lifetime of the metastable state is about 2.2 hours \citep{Drake1971}. The hydrodynamic escape timescale, estimated as the time required for gas to flow from the planetary surface to the Hill radius ($r_{\rm H} \approx 2.5 R_p$), is approx. 5.7 hours, assuming a sound speed of 11~\kms. Other depopulation channels, such as collisional de-excitation by electrons, are included in our post-processing but are much slower. Importantly, ionizations and recombinations occur continuously throughout the gas, and the metastable state reflects a statistical population of many atoms rather than the history of any single one. These timescale comparisons demonstrate that while ionization equilibrium is not strictly assumed, the metastable population is maintained under quasi-steady-state conditions captured by our non-equilibrium post-processing approach.

We also compared temperature profiles along a representative ray near the planet from both \texttt{Athena++} and \texttt{Cloudy}. \texttt{Athena++} assumes a nearly isothermal wind, while \texttt{Cloudy}’s predicted temperature varies with the density profile. Although their temperature structures differ on large scales, the resulting spectra agree to within ~1\%, with \texttt{Cloudy} showing slightly stronger absorption due to marginally higher temperatures near the planet. Achieving self-consistency between photoionization heating and dynamics would require iterative coupling, as implemented in the 1D \texttt{sunbather} code \citep{linssen_open-source_2024}, which links a Parker wind solver with \texttt{Cloudy}. Notably, \texttt{sunbather} yields base wind temperatures ($\approx8500$~K) similar to \texttt{Athena++}. We adopt the \texttt{Athena++} temperature profiles to maintain consistency with the hydrodynamic structure.

\subsection{Simulation results}
We modeled the atmospheric escape of WASP-52~b hydrodynamically in 3D using \texttt{Athena++} as described in Section \ref{ch4:sec:Athena_methods}. As a first step, we wanted to identify a simulation setup that reproduces the mid-transit helium absorption observations reported by \cite{kirk_kecknirspec_2022}. Once this was established, we extended the analysis to explore additional out-of-transit phases, providing insights into the potential morphology of escaping tails or streams, analogous to our observational strategy (see Fig.~\ref{ch4:fig:sketch}). To generate the synthetic spectra based on these 3D simulation snapshots, we used the radiative transfer framework described in Section \ref{ch4:sec:postprocessing_Cloudy}. 

In our observational analysis (Sect.~\ref{ch4:sec:Results_observations}), we focused on the line core of the metastable helium triplet, extracting the equivalent width of the line core (see Fig.~\ref{ch4:fig:Helium_obs}, bottom panel). To compare our models to observations, we calibrated the synthetic spectra generated from the 3D hydrodynamic simulations to match the mid-transit line core EW observed by \cite{kirk_kecknirspec_2022}.

To achieve this calibration, we rescaled the densities and pressures of the simulation snapshots, using a scaling factor $S$. The hydrodynamic equations that govern the outflow are scale-invariant, which means that we can adjust the pressure and density after the simulation without affecting the overall morphology of the wind-wind interactions. Crucially, this approach preserves the ratio of mass-loss rates between the planetary and stellar winds, ensuring the relative dynamics remain consistent. This rescaling allows us to match the simulated line core EWs to the observed mid-transit values reported by \cite{kirk_kecknirspec_2022}, without changing the fundamental properties of the outflow. The scaling factors we applied, along with the resulting time-averaged mass-loss rates $\langle \dot{m} \rangle$ for the star and planet, are summarized in Table~\ref{ch4:tab:mass-loss_rates}.

As a starting point, we explored planetary hydrodynamic escape parameters $\lambda_p$ between 3 and 9. For a direct comparison with the observations of \cite{kirk_kecknirspec_2022}, we convolved the mid-transit synthetic spectra with a Gaussian to match the Keck/NIRSPEC resolution ($R = 25~000$). The results are shown in Appendix \ref{ch4:app:in_transit}. A low escape parameter ($\lambda_p = 3$), corresponding to a hot planetary wind of $T \approx 15~700~$~K, produces a highly thermally broadened line profile that is inconsistent with the observed line width. Conversely, a high escape parameter ($\lambda_p = 9$), representing a cold planetary wind of $T \approx 5200$~K, results in narrow line profiles that do not match the observed broadening. Unlike our previous study \citep{nail_cold_2024}, we do not find kinematic broadening of the helium line for our WASP-52~b simulations. This broadening typically occurs when the line forms in a high-density outflow, a scenario inconsistent with the modest excess helium absorption observed for WASP-52~b. These findings rule out both low and high $\lambda_p$ scenarios.

The best agreement with the observed line width is achieved for $\lambda_p = 5$, corresponding to a planetary wind temperature of approximately $T \approx 9400~$K. Building on this result, we explored four models: one assuming isotropic winds for both the planet and the star (ISO), and three variations involving adjustments to the stellar wind strength (TORUS, TAIL) and the inclusion of a day-night anisotropy in the planetary wind (ANISO).

\begin{table}
    \centering
    \caption{Density scaling factors and time-averaged mass-loss rates from 3D simulation models.}
    \begin{tabular}{lccc}
        \toprule
        \toprule
         model  & $S$ & $\langle \dot{m}_p \rangle $ &  $\langle \dot{m}_* \rangle$ \\ 
          & & [g~s$^{-1}$] & [g~s$^{-1}$]\\ 
         \midrule
         ISO & 1.40 & $2.03 \times 10^{11}$   & $9.98 \times 10^{11}$\\
         ANISO & 0.86 & $1.73 \times 10^{11}$ & $5.92 \times 10^{11}$ \\
         TORUS & 1.20 & $1.36 \times 10^{11}$ & $6.85 \times 10^{10}$\\
         TAIL & 1.40 & $1.74 \times 10^{11}$  & $2.79 \times 10^{12}$\\
        \bottomrule
    \end{tabular}
    \label{ch4:tab:mass-loss_rates}
\end{table}

\begin{figure}[htp!]
    \centering
    \includegraphics[width=.38\textwidth]{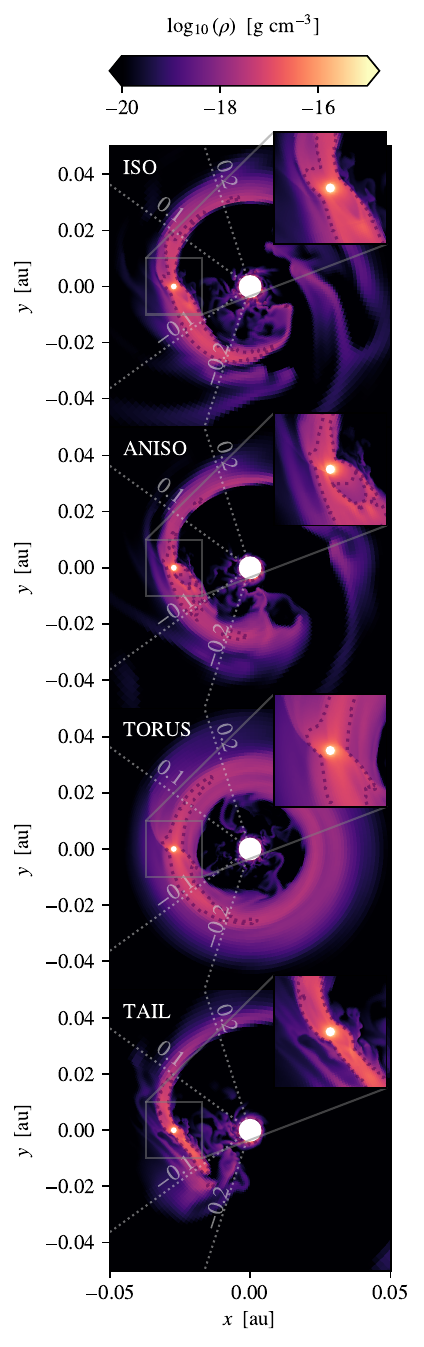}
    \caption{Density distribution of the planetary wind in the orbital mid-plane from the \texttt{Athena++} 3D hydrodynamic simulations. The star is located at the center, with the planet on the left orbiting in the counterclockwise direction. The diagonal dotted lines represent the observing angles of various orbital phases $\varphi$. The dark dotted contours indicate surfaces with densities of $\rho = 1\times 10^{-18}$~g~cm$^{-3}$.}
    \label{ch4:fig:snapshot_grid}
\end{figure}

Figure \ref{ch4:fig:snapshot_grid} presents the scaled snapshots of the four simulation models, captured after five orbits, by which time a quasi-steady state had been reached. In the ISO model, the planetary wind extends up to an orbital phase of approximately $-0.2$, suggesting that it could be probed during pre-transit observations. Near the planet, the outflow shows characteristics intermediate between a bubble and stream morphology, as described by \cite{macleod_streams_2024}. A cavity of unshocked gas is still present close to the planet, but it is stretched along the orbit by tidal forces.

In the ANISO model, the cavity on the planet's night and trailing sides disappears due to the reduced gas pressure on the nightside, which is insufficient to counteract the stellar wind. Consequently, the gas in these regions becomes completely shocked. On the planet’s dayside, however, the ram pressure of the planetary outflow is high enough to enable the formation of an extended cavity that is accelerated along the leading direction of the planet's orbit.

The TORUS model represents a scenario where the stellar wind is too weak to confine the planetary outflow. As a result, the planetary wind escapes freely and is not significantly deflected or compressed. The gas lingers in the planet's orbital path, forming a torus-like structure around the star. In contrast, the TAIL model shows a scenario where the planetary wind is highly confined by a strong stellar wind, resulting in the formation of a dense, trailing tail and the truncation of the leading arm.

\subsection{Synthetic helium results}
Using the snapshots described above, we generated synthetic helium spectra for different observing angles. The resulting line profiles, shown alongside the observational mid-transit data from \cite{kirk_kecknirspec_2022} in Figure \ref{ch4:fig:app_Keck_profiles}, allow for a direct comparison. 

\begin{figure}
        \centering
        \includegraphics[width=0.5\textwidth]{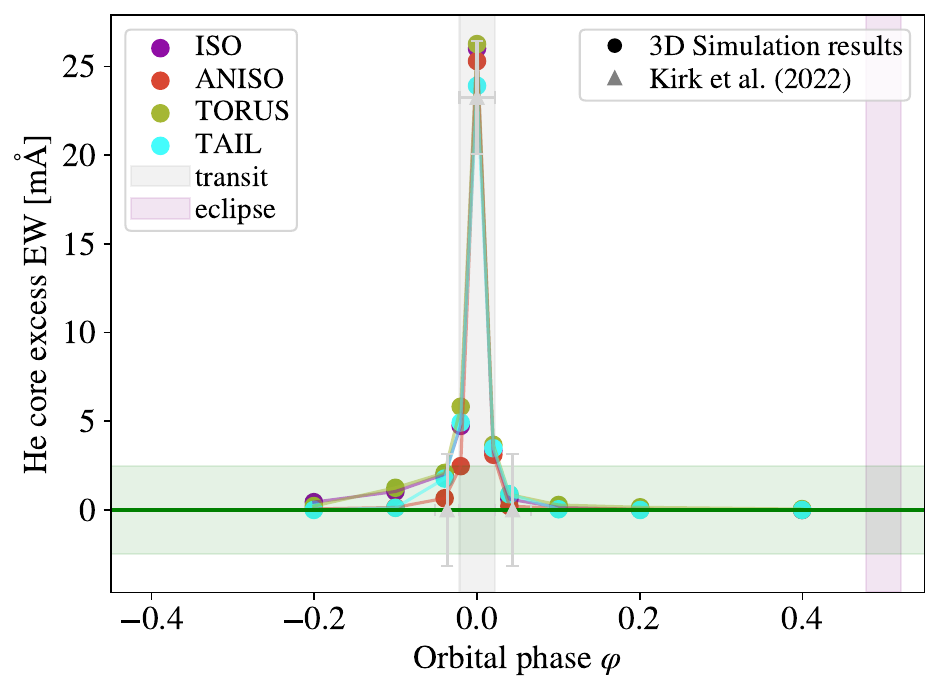}
        \caption{Excess absorption in the helium line core derived from synthetic simulation spectra. Equivalent widths (EWs) were calculated by integrating the flux over the interval $\lambda = [10832.9, 10833.62]~\text{\AA}$, consistent with the observational methodology described in Section \ref{ch4:sec:Results_observations}. The green region represents the observational results from this study, while the gray points indicate the Keck/NIRSPEC data reported by \cite{kirk_kecknirspec_2022} for comparison.}
        \label{ch4:fig:Simulation_EW}
\end{figure}

In this section, however, we focus on the helium line core equivalent width (EW), following the approach used in the observational data analysis in Section \ref{ch4:sec:Results_observations}. To do so, we integrate over the core of the spectral line within the same wavelength range specified in the observational study. The selected orbital phases for reference include $\pm0.2$, $\pm0.1$, and 0. Additionally, we examine ingress and egress phases at $\varphi = \pm0.02$, as well as $\varphi = \pm0.04$, corresponding to the immediate pre- and post-transit points, just before the first contact point ($t_1$) and after the fourth contact point ($t_4$). Although our observations did not cover these additional phases, analyzing them helps constrain the density gradients near the planet predicted by different models and their impact on the expected EW light curves. The integrated EW results for the four models are shown in Figure \ref{ch4:fig:Simulation_EW}. We convolved the synthetic spectra with the instrument resolution of Keck/NIRSPEC (R = 25~000) to match the simulation density scaling with the observational results. Therefore, the synthetic line-core EW presented in Figure \ref{ch4:fig:Simulation_EW} are slightly higher than the mid-transit observational data point, as the figure displays the results of the unconvolved spectra. This figure is intended as a comparison to the observations conducted in this study, and given the high resolving power of the spectrographs we used, the impact on the line profile is negligible.

The EW light curve shows that for all models, no significant out-of-transit absorption in the \hei line core would be apparent in our observations. This means that even if there is a torus of gas lingering in the orbit of the planet, as shown in the TORUS model, no excess absorption is expected to be detected, considering the sensitivity of our observations. Therefore, we cannot definitively rule out any of the scenarios presented here. In all models, except the ANISO model, enhanced helium absorption could be detected during the transit contact point $t_4$ in the trailing direction. For the ANISO model, the night-side pressure is too low to form a tail, and the velocity shift appears too high (see Fig.~\ref{ch4:fig:app_Keck_profiles}). This suggests that a strong day-night anisotropy is unlikely for WASP-52~b, which may explain the lack of a blueshift in the \hei line during transit.

\section{Discussion} \label{ch4:sec:discussion}
\subsection{Outflow morphology of WASP-52b}
Our \hei line observations suggest that the atmospheric outflow of WASP-52~b is not detectably extended. We constrain its extent to $< 73~R_p$ ($\varphi = 0.26 $) in the trailing and to $< 31~R_p$ ($\varphi = -0.11 $) in the leading direction. 
Comparisons with 3D models suggest that the observed in-transit absorption line width is best explained by thermal broadening at a temperature of $T\approx 9400~$K, corresponding to a moderate hydrodynamic escape parameter of $\lambda_p = 5$.

In our previous work \citep{nail_cold_2024}, which explores the outflow morphology of HAT-P-67~b and HAT-P-32~b, we examined the kinematic broadening of the helium line in a stream-like morphology, where the outflow diverges into two cold streams, corresponding to a high $\lambda_p\approx9$ value and a low temperature of $T\approx 4700$~K at the dayside of the planet. This results in a broad and irregular helium line profile, with kinematic broadening dominating over thermal broadening. However, the in-transit spectra of WASP-52~b do not show kinematic broadening, suggesting that the outflow likely does not follow a stream-like morphology. 

Certain limitations have affected the scope of our analysis. One challenge was the lack of NIR observations at phase 0.1 due to a readout error in CARMENES and unfavorable weather conditions during the \cplus observations. This phase would have been particularly valuable for probing the presence of a trailing tail of escaping gas, as suggested by JWST/NIRISS observations from \cite{fournier-tondreau_transmission_2024}. They report a tentative helium detection at 2.9$\sigma$ during egress, which could be attributed to a trailing tail of escaping gas. Although their observations were taken at a much lower resolution (R $\approx$ 700), making direct comparison and translation to the high-resolution spectra presented in this work challenging, their results are broadly consistent with the TAIL, TORUS, and ISO models presented in this study. For a direct comparison, see Figure~\ref{app:fig:full_EW} in Appendix~\ref{app:full_EW}. Specifically, both egress and ingress observations with JWST show enhanced helium absorption, with egress absorption being slightly stronger. The ANISO model, however, can be ruled out for two reasons: first, the light curve does not show enhanced absorption during egress (see Fig.~\ref{ch4:fig:Simulation_EW}), and second, the mid-transit velocity shift of the helium line appears too high (-4.5 \kms, see App.~\ref{ch4:app:in_transit}), whereas \cite{kirk_kecknirspec_2022} reported a shift of $0.00 \pm 1.19$~\kms.

Although WASP-52~b and HAT-P-32~b have similar Hill radii and surface gravities, leading to comparable Rossby numbers, important for distinguishing between stream and bubble morphologies as predicted by \citet{macleod_streams_2024} (see Fig.~9 therein), the helium outflow from WASP-52~b appears significantly less extended. While the results from \cite{nail_cold_2024} suggest that HAT-P-32~b has a stream-like outflow (high $\lambda_p$), this raises the question of what differentiates these two planets in terms of atmospheric escape, or what causes the variation in their escape parameters. One possible explanation lies in differences in optical depth, which may be influenced by a higher planetary mass-loss rate as well as variations in the population of metastable helium. Alternatively, assuming similar tidal and mass-loss conditions, as well as composition, differences in outflow temperature could also play a role. Under this assumption, the outflow temperature of WASP-52~b seems to be higher (lower $\lambda_p$) than that of HAT-P-32~b. 

One possibility is that a different wind-driving mechanism is at play. According to \citet{owen_mapping_2024}, the Bondi radius (the radius at which bolometrically heated gas becomes unbound) lies below the altitude of the XUV-heated region if core-powered mass-loss is dominant. The altitude of this XUV-heated region, depends strongly on atmospheric composition and the stellar XUV flux. Given that WASP-52 is highly active, it may produce a stronger XUV flux than HAT-P-32, leading to a deeper XUV penetration depth and, as a result, a hotter, photoevaporation-driven outflow. Although our understanding of heat redistribution and gas circulation at very low pressures in the upper atmosphere is limited, one could speculate that heat at the wind launching altitude might be distributed efficiently. If heat redistribution were less effective, we would expect a strong day-to-nightside temperature contrast in the photoevaporation scenario. However, our models show that a pronounced day-night anisotropy (ANISO) seems unlikely, suggesting that other factors, such as atmospheric composition or thermal structure, may play a more significant role in shaping the outflow.

We suggest that the new radiative transfer framework introduced in this work may help resolve some of these uncertainties, as it allows us to calculate the number densities of multiple species simultaneously in a 3D outflow. This capability opens up the possibility of identifying potential tracers of atmospheric escape and the composition of the outflow. Different species probe different line-forming regions: lower altitudes, where the wind is launched, and higher altitudes, where the wind interacts with the stellar wind. Helium, for instance, serves as a good probe of the upper atmosphere due to atmospheric fractionation, while heavier species could be probing the lower atmosphere. 

An important consideration in the study of WASP-52~b is the distinction between the physical extent of its outflow and its detectable extent. While simulations indicate the presence of an extended atmosphere, the observed signal depends on the outflow’s geometry, optical depth, and level populations. The densities in the extended outflow may be too low to effectively shield ionizing radiation, potentially contributing to the production of doubly ionized helium. Additionally, recombination is sensitive to gas density; in a low-density environment, He$^{2+}$ ions encounter electrons less frequently, reducing the likelihood of recombination into He$^+$. Consequently, while the material is present, its transparency in the observed helium line could explain the weak or absent signal during out-of-transit observations. This highlights the importance of considering not just the presence of an outflow, but also its radiative properties. Alternative tracers that are more sensitive to conditions in lower-density regions, such as ionized species that remain detectable even at lower column densities, may offer additional insight into the structure and composition of the outflow.

\subsection{Challenges from stellar activity in detecting extended atmospheric features}

Stellar activity remains a potential complication when studying WASP-52~b as a test case for atmospheric escape. Previous observations of the planet have shown that stellar activity can interfere with the interpretation of atmospheric escape tracers \citep[e.g.][]{bruno_wasp-52b_2020, chen_detection_2020, fournier-tondreau_transmission_2024}, as atomic lines observed in the planetary atmosphere are also chromospherically sensitive lines. \citet{Cauley_effects_2018} demonstrated that heterogeneities on the stellar surface can introduce spurious in-transit absorption signals in high-resolution transmission spectra. Their simulations showed that unocculted active regions can produce apparent absorption depths of up to 0.3\% in H$\alpha$ and 0.2\% in the Na\,{\sc i} D lines. By contrast, they found that the \hei line is largely unaffected by stellar activity; in many cases, activity tends to dilute rather than mimic the absorption signal. This, of course, strongly depends on the location of the active regions relative to the transit chord. If active regions are predominantly unocculted, they can weaken the contrast between the stellar and planetary signals, whereas occulted active regions may still introduce subtle distortions. Nonetheless, compared to other chromospheric lines like \ha, the \hei triplet seems to be one of the most reliable tracers for atmospheric escape in active stars. This further motivated our focus on this line in our study. 

Given the potential impact of stellar activity on transmission spectra, one approach to disentangle stellar and planetary contributions is to monitor both H$\alpha$ and He\,{\sc i} simultaneously. Investigating a potential anticorrelation between the two lines could indicate stellar activity \citep{guilluy_gaps_2020}. This highlights another advantage of our new radiative transfer framework, which can help quantify the planetary contribution of atmospheric escape tracers that also probe stellar activity. However, a limitation of the present study is the lack of spectroscopic coverage during the immediate pre- and post-transit phases. These orbital phases are particularly valuable for identifying extended or asymmetric features, such as trailing tails or leading arms of escaping material. In an ideal future experiment, these phases would be sampled more densely, with higher cadence observations in both the near-infrared and visible wavelength ranges. Obtaining at least five or more data points outside of transit, would improve our ability to track variability in the line profiles and to test whether \ha and \hei signals are anti-correlated. This would help to separate stellar and planetary contributions in the tracer lines.

To better constrain the baseline variability of the stellar lines, a long-term monitoring campaign during the secondary eclipse, or at other out-of-transit phases, would also be valuable. Such monitoring, carried out over multiple nights, could characterize the intrinsic variability of stellar chromospheric lines and provide the context needed to confidently interpret in-transit absorption features.

\section{Summary} \label{ch4:sec:summary}
In this study, we present high-resolution observations of the helium triplet at 10833~\AA\ using the spectrographs \cplus and CARMENES, covering an extended range of orbital phases ($\varphi \approx \pm 0.1, \pm 0.2, 0.5$) to investigate the extent of the planet's helium outflow. To explore possible outflow morphologies, we performed 3D simulations with \texttt{Athena++}, testing four distinct scenarios and comparing them to the observational data. Additionally, we used an improved radiative transfer post-processing approach, extending our previous methods to generate synthetic spectra, which could potentially be used to model other atmospheric escape tracer lines. Our results on the atmospheric escape of WASP-52~b include:  
\begin{itemize}
    \item[$\bullet$] Our helium observations constrain the extent of metastable helium in the orbit of WASP-52~b, ranging from a minimum at phase $-0.11$ ($31~R_p$) to a maximum at phase $0.26$ ($73~R_p$) in the trailing direction.
    \item[$\bullet$] We consider a stream-like morphology, as suggested for HAT-P-67~b and HAT-P-32~b, to be unlikely for WASP-52~b. Instead, we propose that its outflow morphology likely represents an intermediate regime between bubble and stream structures. This conclusion is supported by a hydrodynamic escape parameter of approximately $\lambda = 5$, corresponding to an outflow temperaure of $T\approx9400~$K, which is required for our 3D simulations to reproduce in-transit observations by \cite{kirk_kecknirspec_2022}.
    \item[$\bullet$] Assuming a fixed hydrodynamic escape parameter, the extent and shape of the planetary wind, whether it forms a torus or a trailing tail, depend on the stellar mass-loss rate; however, from our modeling perspective, the two scenarios appear indistinguishable, as the metastable helium densities in a potentially extended structure are too low to be detected with our current observations.
    \item[$\bullet$]A high day-night anisotropy scenario is unlikely, as it would produce a blueshift in the helium line during transit, which is not observed by \cite{kirk_kecknirspec_2022}. Furthermore, such a scenario would result in higher helium absorption during ingress than during egress, which contradicts the low-resolution observations with JWST by \cite{fournier-tondreau_transmission_2024}, where egress absorption is observed to be slightly stronger.
    \item[$\bullet$] The new radiative transfer framework presented here, which uses \texttt{Cloudy} to calculate the number densities of multiple species in the 3D simulations of wind-wind interaction between the star and planet, has the potential to model multiple atmospheric escape tracer lines and provide more robust results in low-temperature and low-density regimes. When combined with observations, it offers great potential for accessing the planetary wind structure in terms of composition and temperature.
\end{itemize}

\section{Data availability}
The observational data, simulation snapshots, and Jupyter notebook for reproducing the figures\footnote{\url{https://doi.org/10.5281/zenodo.16687147}}, along with the new radiative transfer framework and tutorial notebooks for running Athena++ simulations and Cloudy post-processing\footnote{\url{https://doi.org/10.5281/zenodo.16687969}}, are all publicly available on Zenodo.

\begin{acknowledgements}
Based on observations made with ESO Telescopes at the La Silla Paranal Observatory under program ID 112.25M6.001. 
This project has received funding from the European Union's Horizon 2020 research and innovation programme under grant agreement No 101004719.
We thank SURF (\url{www.surf.nl}) for the support in using the National Supercomputer Snellius.
A. Oklop\v{c}i\'{c} gratefully acknowledges support from the Dutch Research Council NWO Veni grant.
\end{acknowledgements}

\bibliography{lib_paper}{}
\bibliographystyle{aa}

\begin{appendix}

\section{Full helium line analysis} \label{app:full_EW}
\begin{figure}[h!]
    \centering
    \includegraphics[width=.5\textwidth]{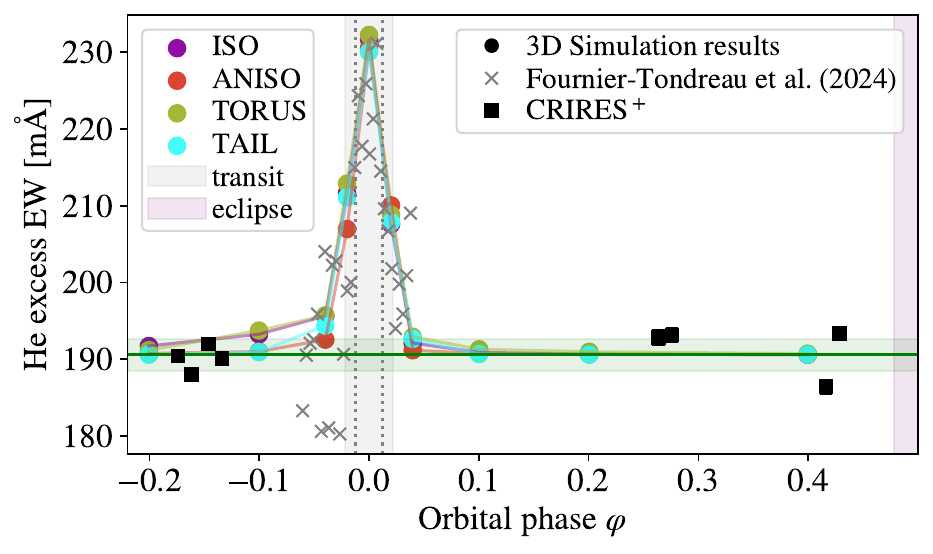}
    \includegraphics[height=0.22\textheight]{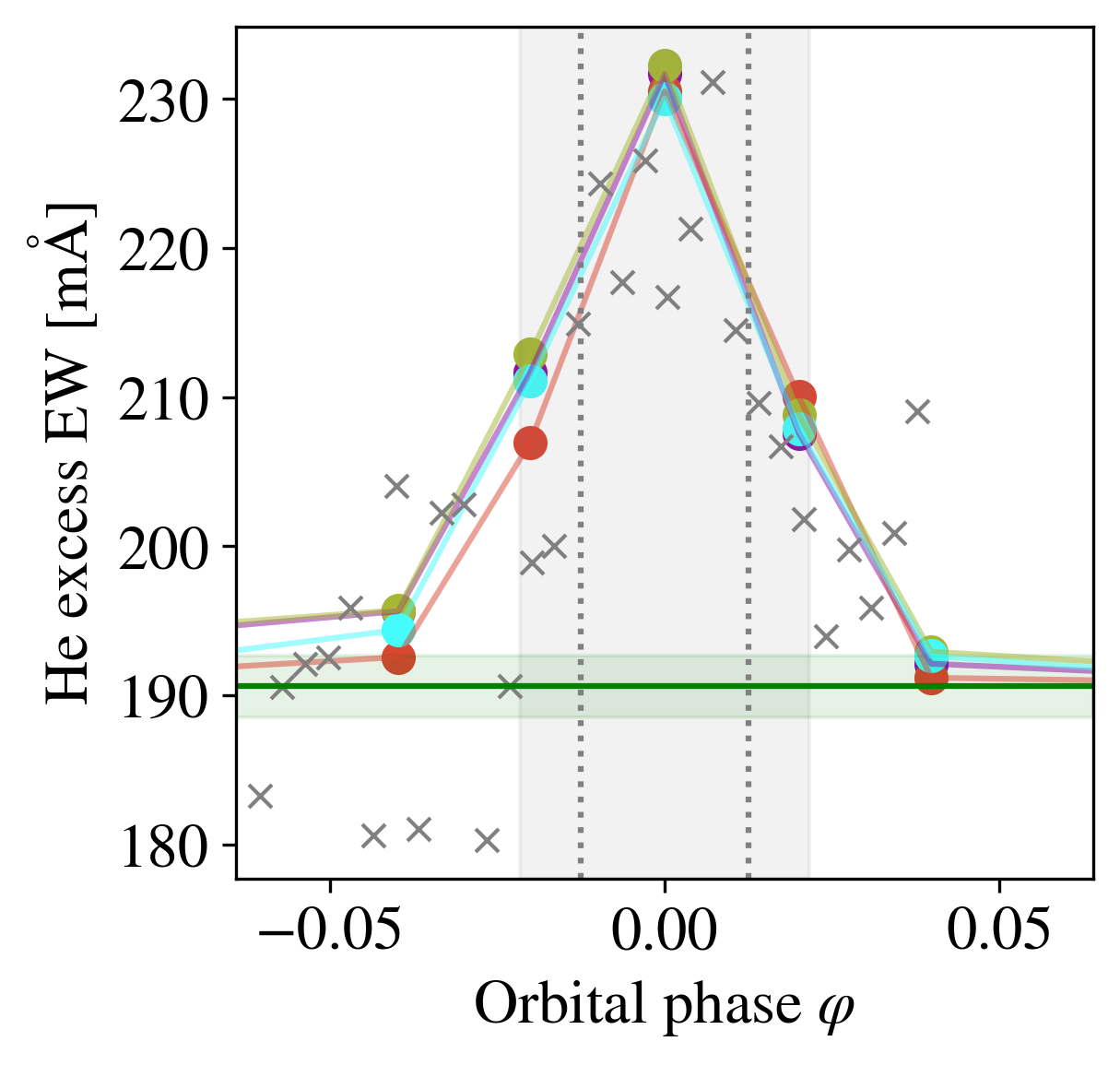}
    \caption{Full-line equivalent width measurements derived from \cplus\ data and the four synthetic models presented in this study. A constant offset was added to the model curves to align them with the mean of the observed CRIRES+ values, which are interpreted as the stellar baseline. For comparison, we also include JWST/NIRISS data from \citet{fournier-tondreau_transmission_2024}; the flux values were digitized and converted to equivalent width assuming a Gaussian instrumental profile with spectral resolution $R \approx 700$. The estimated uncertainty in the JWST-derived EW values is $\sim$15.5~m\AA, but error bars are omitted here for clarity. The bottom panel shows a zoom-in of the region around the transit for better visibility, highlighting the details of the absorption features during ingress and egress, indicated by vertical dotted lines}
    \label{app:fig:full_EW}
\end{figure}

Figure \ref{app:fig:full_EW} shows the equivalent width (EW) measurements derived from a full-line integration over the helium triplet feature ($\lambda = 10831.556 - 10834.300$~\AA), using only the \cplus\ data. The \cplus\ spectra enable such an analysis, as they provide complete coverage of the line profile without contamination from telluric emission lines. For consistency, we applied the same integration procedure to the four simulation models presented in this study (see Tab.~\ref{ch4:tab:model_overview}). The \cplus\ EW measurements cluster around a nearly constant value of $190.6 \pm 2.1$~m$\AA$, which we interpret as the contribution from the stellar line alone. To compare the observational and simulated results, we added this stellar baseline offset to the model EWs.

We also include data points extracted from the JWST/NIRISS observations published by \citet{fournier-tondreau_transmission_2024}, which offer higher time resolution but considerably lower spectral resolution ($R \approx 700$) compared to the KECK/NIRSPEC data presented by \citet{kirk_kecknirspec_2022} (see Fig.~\ref{ch4:fig:Helium_obs} and \ref{ch4:fig:Simulation_EW}). To enable a direct comparison with the high-resolution analysis presented in this work, we digitized the flux values (reported in ppm) from their Figure 11 using \texttt{WebPlotDigitizer}\footnote{\url{https://github.com/automeris-io/WebPlotDigitizer}}, and converted them into fractional excess absorption. To derive the equivalent width (EW), we first converted the absorption values from parts per million (ppm) to unitless fractional values by dividing by $10^6$. Assuming a Gaussian instrumental profile, the equivalent width was then computed using the following relation:
\begin{equation}
\mathrm{EW} = f \cdot \Delta\lambda = f \cdot \frac{\lambda_0}{R},
\end{equation}
where $f$ is the fractional excess absorption, $\lambda_0$ is the central wavelength of the helium feature, and $R$ is the spectral resolution of the instrument. The quantity $\Delta\lambda = \lambda_0 / R$ corresponds to the effective width of the resolution element or kernel area. For the JWST/NIRISS data, we adopted $\lambda_0 = 10830$~\AA\, which is the central wavelength reported in their analysis, and $R = 700$. The uncertainty in the flux measurements is approximately 1000~ppm, which corresponds to an equivalent width error of 15.5~m\AA. For clarity of presentation, we omit these error bars in Figure~\ref{app:fig:full_EW}.

One advantage of using the full wavelength range of the line, is that it captures planetary signals that are shifted due to the planet's motion. This approach is also more robust to data sensitivity, as it integrates over a broader wavelength range. By using the higher time-resolution JWST data for comparison, we obtain a better sampling of the ingress and egress phases, providing valuable insights into the immediate extent of the atmosphere. The JWST data suggest that egress and post-transit absorption are slightly enhanced compared to ingress and pre-transit, which could indicate a trailing, escaping atmosphere. However, this conclusion is tentative and requires further investigation.

\section{Convergence testing of the new radiative transfer framework} \label{app:convergence}
To assess the numerical convergence of the new radiative transfer framework, described in Section \ref{ch4:sec:postprocessing_Cloudy}, we tested its sensitivity to the number of rays used in the spectrum calculation. Specifically, we computed the helium absorption spectrum for the TORUS model while varying the total number of rays. In this test, we distinguish between regions of differing density: a high-density region, probed during mid-transit observations at phase $\varphi=0$, and lower density region, probed for example during pre-transit observations, here at phase $\varphi=-0.024$. 

\begin{figure}[h!]
    \centering
    \includegraphics[width=0.46\textwidth]{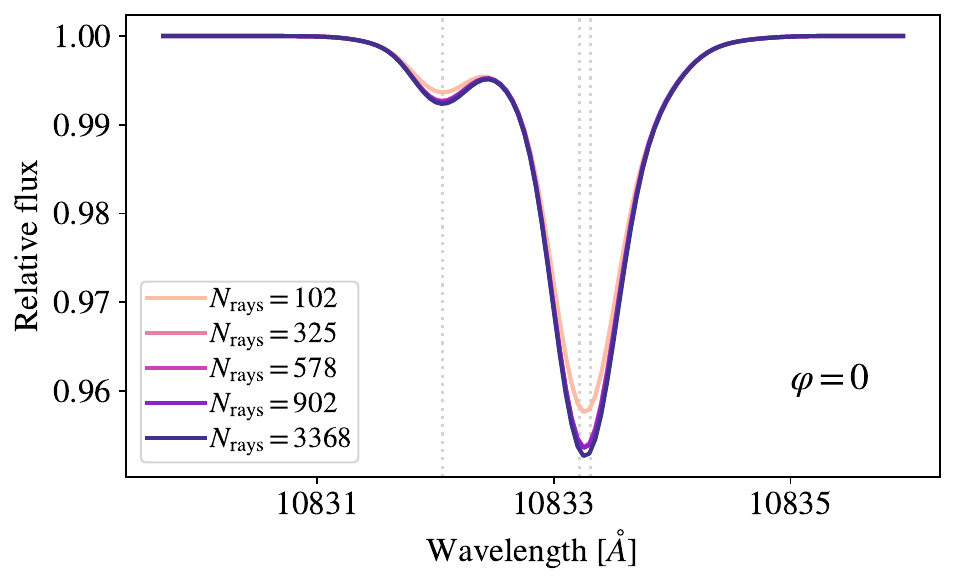}
    \includegraphics[width=0.46\textwidth]{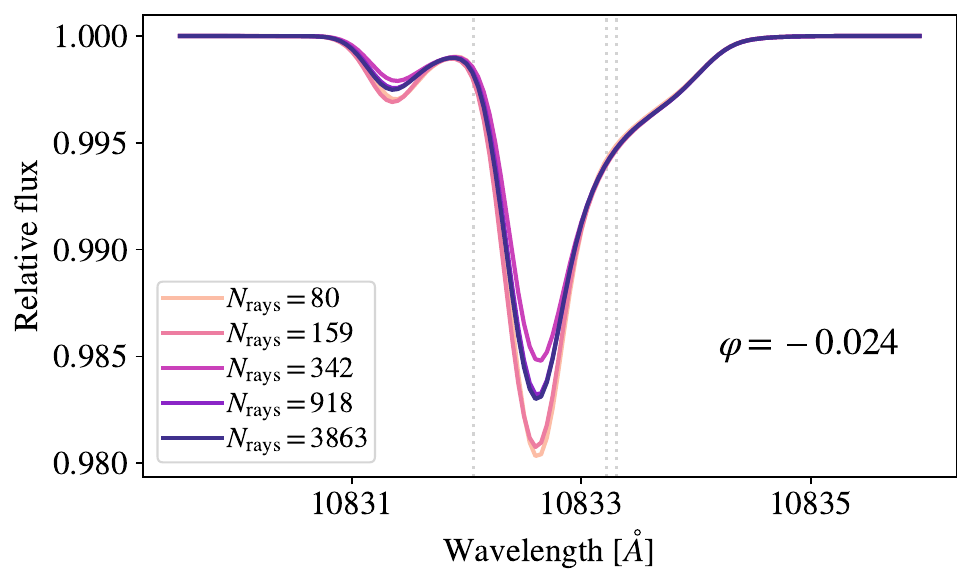}
    \caption{Convergence of the helium absorption spectrum with respect to the number of rays used in the RT calculations. The top panel shows the in-transit spectrum at orbital phase $\varphi = 0$, while the bottom panel shows the pre-transit spectrum at $\varphi = -0.024$. Light gray vertical dotted lines mark the nominal wavelengths of the \hei triplet. In the in-transit case, the spectrum converges quickly, with little change beyond $N_\mathrm{rays} > 300$. However, the pre-transit configuration, which probes more extended and lower-density regions, requires a higher number of rays to achieve convergence. For this study, all spectra were computed using $N_\mathrm{rays} \approx 1000$.} 
    \label{fig:ray_convergence}
\end{figure}

Figure~\ref{fig:ray_convergence} shows the resulting helium absorption spectra computed with varying numbers of rays. Spectra calculated with approximately 100 to 300 rays show noticeable fluctuations, particularly in low-density regions. However, the spectra converge rapidly as the ray count increases, with stable results achieved around 900 rays. While a slight difference in the spectrum is observed between 1000 and 3000 rays, the additional computational cost does not yield a significant improvement in accuracy. This indicates that using approximately 1000 rays per model snapshot provides sufficient convergence for the cases examined in this study. 

\section{Comparison of radiative transfer methods} \label{ch4:app:old_new_RT_compare}

The previous radiative transfer (RT) code, first introduced in \cite{macleod_stellar_2022} and further used in \cite{nail_cold_2024} and \cite{nail_cold_2024}, was built for high-temperature regimes of $10^4$~K. However, planetary winds formed under lower hydrodynamic escape parameters, such as those seen in stream-morphology, have significantly lower temperatures on the order of $10^3$~K. Therefore, we were motivated to improve our RT method by incorporating the more sophisticated NLTE plasma simulation code \texttt{Cloudy} (see Sect.~\ref{ch4:sec:postprocessing_Cloudy}), that gives access to a more complex chemical network and that reacts more appropriately to different temperature regimes.

\begin{figure}
    \centering
        \includegraphics[width=.46\textwidth]{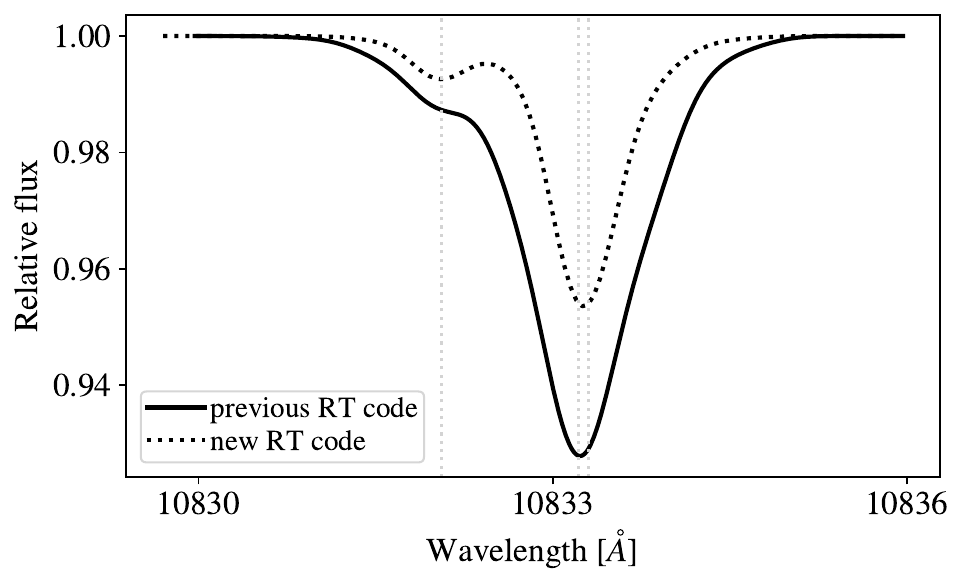}
        \includegraphics[width=.46\textwidth]{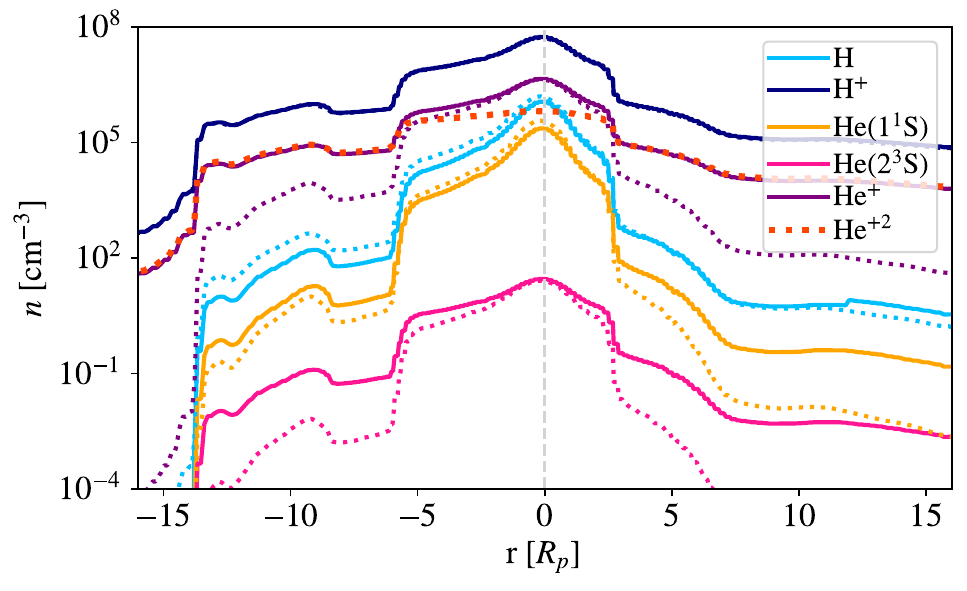}
        \vfill
        \includegraphics[width=.46\textwidth]{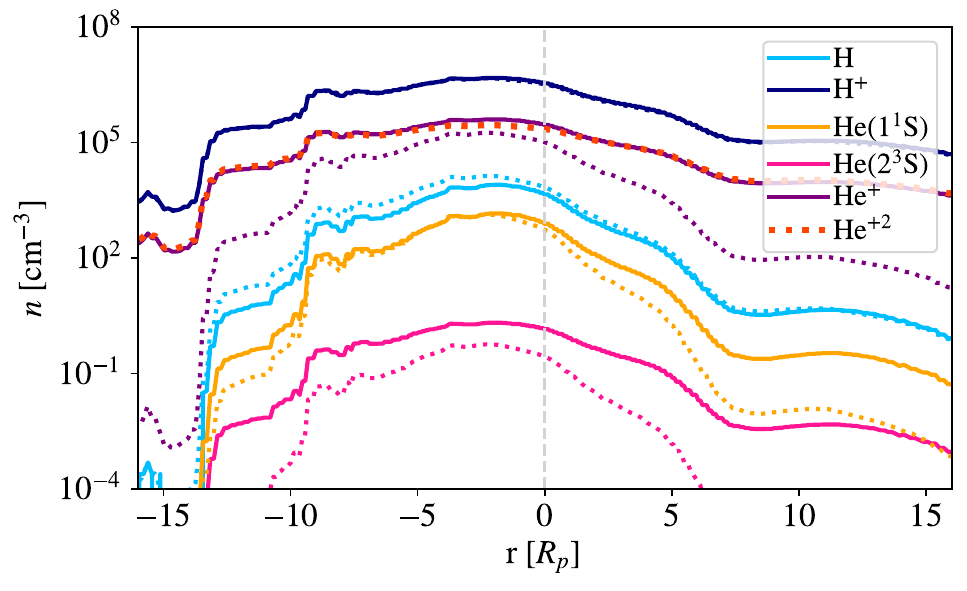}
        \caption{Comparison of the helium spectrum (top) and the number densities of hydrogen and helium species (middle and bottom) between the previous and new radiative transfer (RT) methods for the TORUS model. Dotted profiles show results from the new RT code used in this study, which includes \texttt{Cloudy}, while solid lines represent the earlier code developed by \citet{macleod_stellar_2022}, based on the radiative transfer formalism introduced in \citet{oklopcic_new_2018}. Light gray vertical dotted lines mark the nominal wavelengths of the \hei triplet. The middle  panel shows a ray crossing near the planet $\mathrm{YZ}(0.16|0.02)R_*$, while the bottom panel shows a more distant ray $\mathrm{YZ}(0.86|-0.28)R_*$, both in a mid-transit configuration $(\varphi=0)$. The previous code lacks He$^{2+}$ formation, which becomes significant in low-density regimes, reducing He$^{+}$ available for recombination into the metastable He(2$^3$S) state that forms the observed spectral lines.}
    \label{ch4:fig:rt_comparison}
\end{figure}

To further evaluate the differences between the previous and new RT methods, we compared the spectra and species number densities from a snapshot of the TORUS model, as presented in Section \ref{ch4:sec:Results_3DSimulations}. The comparison is shown in Figure~\ref{ch4:fig:rt_comparison}, which illustrates the helium spectrum resulting from the radiative decay of the metastable helium state (top panel), along with the number densities of hydrogen and helium species (middle and bottom panel) for both RT methods. While \texttt{Cloudy} provides access to an even broader range of species, we focus here on those included in the previous code.

The dotted lines in Figure~\ref{ch4:fig:rt_comparison} indicate results obtained using the new RT code implemented in this study, while the solid lines represent results from the previous RT code. The middle panel displays the number density profiles of a ray that closely crosses the planet in the $yz$-plane (the plane of observation during the planet's transit in front of the star) at $\mathrm{YZ}(0.16|0.02) R_*$, whereas the bottom panel illustrates the number densities for a ray that passes farther from the planet at $\mathrm{YZ}(0.86|-0.28)R_*$, both in a mid-transit configuration $(\varphi=0)$.

While hydrogen abundances are similar in both codes, the helium species show notable differences. In the new RT code, neutral helium abundances are lower, particularly in low-density regions far from the planet. For regions very close to the planet (within a few planetary radii) the number density profiles of both codes are nearly identical, but differences become more pronounced farther out, as seen in the bottom panel. This discrepancy is due to the formation of He$^{2+}$ in the new code, which the previous code does not account for. At distances from the planet greater than $\sim5~R_p$, most helium in the new RT model transitions to He$^{2+}$, reducing the availability of He$^{+}$ for recombination into the metastable He(2$^3$S) state responsible for the spectral lines at 10833~\AA. For a detailed description of helium transitions, we refer to Fig.~1 in \cite{oklopcic_new_2018}.

\begin{figure}
    \centering
    \includegraphics[width=0.46\textwidth]{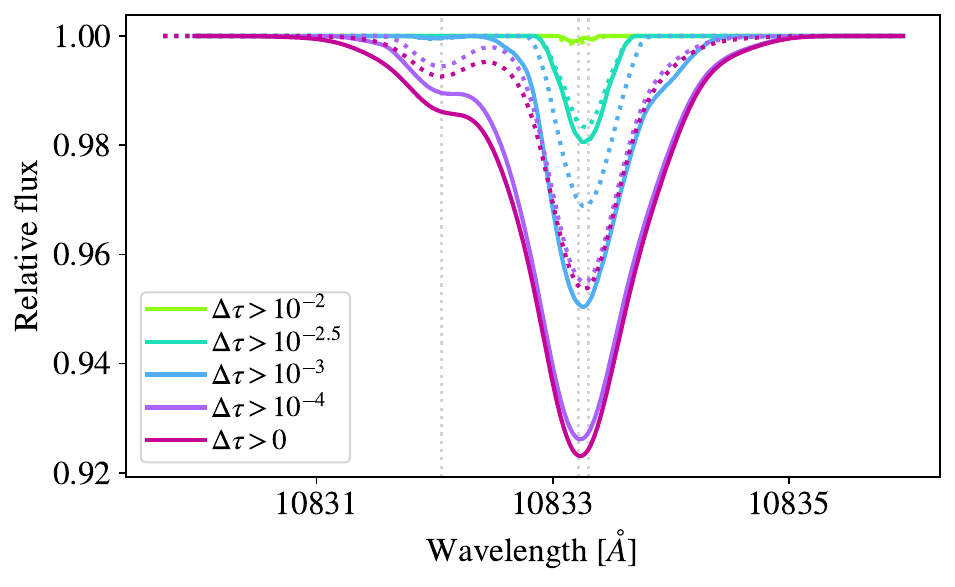}
    \caption{Helium spectra computed using the new (dotted lines) and previous (solid lines) RT codes. Only simulation cells exceeding a specified optical depth threshold in the metastable helium component ($\Delta\tau$) are included. From green to pink, lines progressively incorporate more cells with lower $\Delta\tau$ values. The two codes produce nearly identical spectra when only high-opacity (high-density) cells are included, but begin to diverge as increasingly low-opacity cells are added. The pink line corresponds to the full spectrum shown at the top of Figure~\ref{ch4:fig:rt_comparison}. Light gray vertical dotted lines indicate the nominal wavelengths of the \hei triplet.}
    \label{fig:app:dtau}
\end{figure}

To further illustrate this point, Figure~\ref{fig:app:dtau} compares the helium spectra produced by the new and previous RT codes. In this test, we include only simulation cells that exceed a given optical depth threshold in the metastable helium component ($\Delta\tau$), gradually relaxing the threshold to include more of the low-density material. As expected, both codes yield nearly identical spectra when only high-opacity $\Delta\tau>10^{-2.5}$ (i.e., high-density, close-in) regions are considered. However, as lower-opacity cells, located farther from the planet, are included, the spectra begin to diverge. This behavior supports the idea that the main difference between the two codes arises from the treatment of doubly ionized helium. 

The absence of He$^{2+}$ formation in the previous RT code is acceptable for 1D modeling, which is centered on the planet, but poses a limitation in more complex 3D outflows, in which many light rays do not come close (within a few planet radii) to the planet. This comparison emphasizes the importance of including detailed atomic processes, like He$^{2+}$ formation, to accurately capture low-density outflow regions.

Furthermore, we tested whether the inclusion of metals in \texttt{Cloudy} affects the spectra. However, excluding metals yielded the same spectra as including them, allowing us to narrow the focus to the formation of doubly ionized helium. 

\section{Comparison of 3D model spectra with in-transit observations}\label{ch4:app:in_transit}
\begin{figure}
    \centering
    \includegraphics[width=0.46\textwidth]{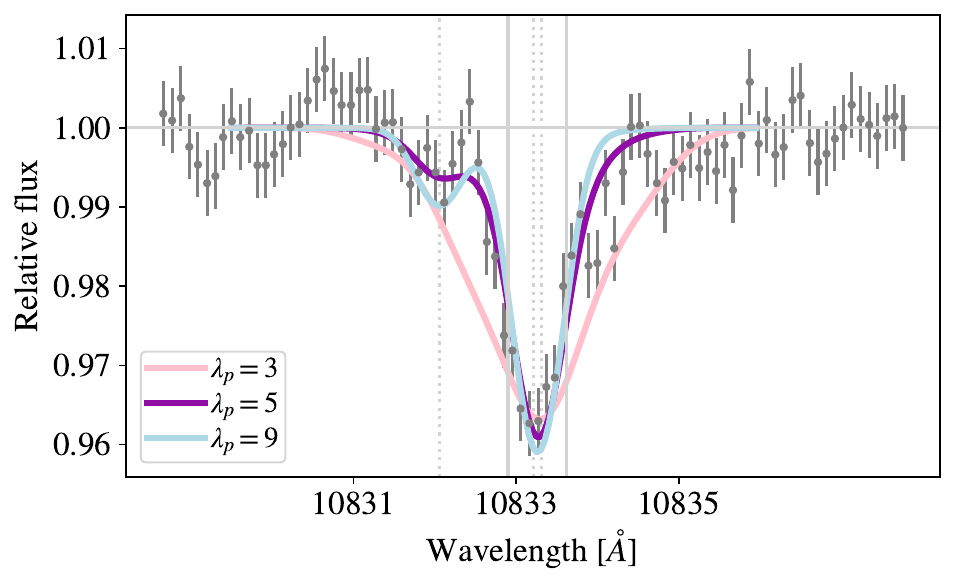}
    \includegraphics[width=0.46\textwidth]{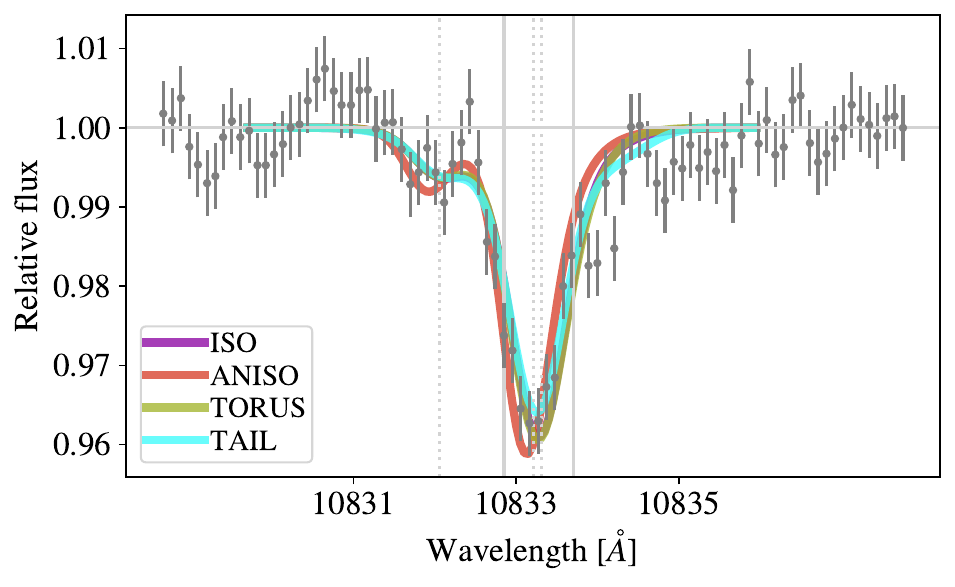}
    \caption{Comparison of synthetic spectra from the four 3D simulation models (colored lines) with in-transit observations of WASP-52~b from Keck/NIRSPEC (dark gray; \citealt{kirk_kecknirspec_2022}). Synthetic spectra are convolved to match the instrument resolution (\( R = 25~000 \)). Light gray dotted lines mark the nominal wavelengths of the \hei triplet; solid lines indicate the line core used for equivalent width (EW) integration. Simulations are scaled to approximately match the observed line width and line core EW. The top panel shows the results for the ISO model with varying hydrodynamic escape parameter \( \lambda_p \): low values (hot winds) produce broader lines, while high values (cool winds) yield lines that are too narrow. An intermediate value (\( \lambda_p = 5 \)) was adopted for all models. The bottom panel shows spectra from all four models of this study using \( \lambda_p = 5\).
}
    \label{ch4:fig:app_Keck_profiles}
\end{figure}

As previously described, we tested various values of $\lambda_p$ for the simulation, which dictates the outflow temperature of the planetary wind (see Eq.~\ref{ch4:eq:T}), and consequently, the width of the line due to thermal broadening (see top panel in Figure~\ref{ch4:fig:app_Keck_profiles}). We chose $\lambda_p = 5$ as the optimal value, which corresponds to an outflow temperature of $T\approx9400$~K.

The bottom panel of Figure \ref{ch4:fig:app_Keck_profiles} presents the results for the four models presented in Section \ref{ch4:sec:Results_3DSimulations}. All models align with the observational data, extracted from Fig.~7 of the publication using \texttt{WebPlotDigitizer}; however, the ANISO model shows a blueshift of $-4.5$ \kms, which does not match the observational data, which was reported as $0.00 \pm 1.19$~\kms. Nevertheless, we wanted to explore this scenario to investigate how it would affect the symmetry of the outflow.

\end{appendix}
\end{document}